\documentclass[mca,article,accept,moreauthors,pdflatex,a4paper]{Definitions/mdpi_arXiv}
\usepackage{hyperref}
\firstpage{1} 
\pubvolume{24}
\issuenum{2}
\articlenumber{56}
\pubyear{2019}
\copyrightyear{2019}
\history{Received: 31 March 2019; Accepted: 24 May 2019; Published: 27 May 2019}

\Title{Finite Strain Homogenization Using a Reduced Basis and Efficient Sampling}

\Author{Oliver Kunc \orcidA{} and Felix Fritzen *\orcidB{}}

\AuthorNames{Oliver Kunc and Felix Fritzen}

\address[1]{%
Efficient Methods for Mechanical Analysis, Institute of Applied Mechanics (CE), University of Stuttgart, 70569 Stuttgart, Germany; kunc@mechbau.uni-stuttgart.de
}
\corres{Correspondence: fritzen@mechbau.uni-stuttgart.de}

\abstract{
The computational homogenization of hyperelastic solids in the geometrically nonlinear context has yet to be treated with sufficient efficiency in order to allow for real-world applications in true multiscale settings.
This problem is addressed by a problem-specific surrogate model founded on a reduced basis approximation of the deformation gradient on the microscale. The setup phase is based upon a snapshot POD on deformation gradient fluctuations, in contrast to the widespread displacement-based approach. In order to reduce the computational offline costs, the space of relevant macroscopic stretch tensors is sampled efficiently by employing the Hencky strain\color{black}.
Numerical results show speed-up factors in the order of 5--100 and significantly improved robustness while retaining good accuracy. An open-source demonstrator tool with 50 lines of code emphasizes the simplicity and efficiency of the method.
}



\newtheorem{corollary}[definition]{Corollary}
\newtheorem*{corollary*}{Corollary}
\newtheorem{note}{Note}
\newtheorem*{note*}{Note}

\newtheorem*{conclusion*}{Conclusion}

\newcommand{\TT}{^{\sf T}}
\newcommand{\RB}{^{\rm RB}}

\newcommand{\olffC}{\ol{\ffC}}
\newcommand{\olW}{\ol{W}}
\newcommand{\olfF}{\ol{\fF}}
\newcommand{\olfP}{\ol{\fP}}

\newcommand{\olfU}{\ol{\fU}}
\newcommand{\olfH}{\ol{\fH}}

\newcommand{\pmo}{\phantom{-}0}

\newcommand{\pd}[2]{\displaystyle\frac{\displaystyle\partial #1}{\displaystyle\partial #2}}
\newcommand{\pdII}[2]{\displaystyle\frac{\displaystyle\partial^2 #1}{\displaystyle\partial {#2}^2}}
\newcommand{\tr}[1]{{\rm tr  }( #1 )}
\newcommand{\lb}{\left(}
\newcommand{\rb}{\right)}
\newcommand{\sv}{\lb\begin{array}{ccccccccccccccccc}}
\newcommand{\sV}{\begin{bmatrix}}
\newcommand{\eV}{\end{bmatrix}}
\newcommand{\ev}{\end{array}\rb}

\newcommand{\sty}[1]{\mbox{\boldmath $#1$}}
\newcommand{\styy}[1]{{\mathbb{#1}}}

\newcommand{\fa}{\sty{ a}}
\newcommand{\fb}{\sty{ b}}

\newcommand{\fe}{\sty{ e}}

\newcommand{\fu}{\sty{ u}}
\newcommand{\fv}{\sty{ v}}
\newcommand{\fw}{\sty{ w}}
\newcommand{\fx}{\sty{ x}}

\newcommand{\fA}{\sty{ A}}
\newcommand{\fB}{\sty{ B}}

\newcommand{\fF}{\sty{ F}}

\newcommand{\fH}{\sty{ H}}
\newcommand{\fI}{\sty{ I}}

\newcommand{\fP}{\sty{ P}}

\newcommand{\fR}{\sty{ R}}

\newcommand{\fU}{\sty{ U}}

\newcommand{\fX}{\sty{ X}}

\newcommand{\fzero}{\sty{ 0}}

\newcommand{\ffC}{\styy{ C}}

\newcommand{\ffI}{\styy{ I}}

\newcommand{\ffR}{\styy{ R}}

\newcommand{\feps}{\mbox{\boldmath $\varepsilon $}}

\newcommand{\cF}{{\cal F}}

\newcommand{\cO}{{\cal O}}

\newcommand{\cU}{{\cal U}}

\newcommand{\bk}{^{( k )}}
\newcommand{\bm}{^{( m )}}
\newcommand{\bn}{^{( n )}}
\newcommand{\bi}{^{( i )}}
\newcommand{\bj}{^{( j )}}
\newcommand{\bp}{^{( p )}}

\newcommand{\ol}[1]{\overline{#1}}
\newcommand{\ull}[1]{\underline{\underline{#1}}}

\makeatletter
\usepackage[ruled,lined]{algorithm2e}
\SetAlgoSkip{bigskip}
\LinesNumbered
\g@addto@macro{\@algocf@init}{\SetKwInOut{Input}{Input}}
\g@addto@macro{\@algocf@init}{\SetKwInOut{Output}{Output}}

\makeatother

\begin{document}

\section{Introduction}
\label{sec:intro}
\subsection{Purpose}
\label{sec:intro:purpose}
The description of solid mechanics under finite strains is of particular interest in both academia and industry. It allows for accurate descriptions of rotations and stretches under mild assumptions. Thus, many geometric effects can be captured. For instance, alignments and rearrangements of the respective structures may trigger pronounced stiffening or softening effects.

In such cases where rotations and deformations are not suitable for linearization, dissipative effects also play a notable role for many materials.
Regardless of the kind of dissipation involved in a certain process, hyperelasticity usually persists to a certain extent. Therefore, it is worthwhile investigating this comparatively simple case at first, before introducing history dependence into the description. Prominent examples of materials that require a hyperelastic description at finite strains include carbon black-filled rubber~\citep{rendek2010} and amorphous glassy polymers~\citep{nguyen2016}, to name just two.

The main purpose of this work is the \emph{computationally efficient quasi-static homogenization of hyperelastic solids with full account for geometric nonlinearities}. The employed methodology is twofold. First, a Reduced Basis (RB) model for the microscopic problem is established.  The term Reduced Basis, used in this work, is not to be confused with the homonymous method introduced by Barrault, Maday, Nguyen, and Patera~\cite{Barrault2004}. \color{black} Once set up, it enables more efficient evaluations of the homogenized material response as compared to the Finite Element Method (FEM). Second, an efficient strategy for sampling of the space of macroscopic kinematic states is proposed. This renders the setup phase of the RB model more rational.

\subsection{State of the Art}
\label{sec:intro:state}
Efficiently determining the overall solid--mechanical properties of microstructures has been investigated for decades, and a large body of literature is available. Comprehensive review articles, such as \cite{geers2016} and \cite{saeb2016}, summarize the progress. Here, attention is restrained to few methods most similar or relevant to the present work.

The FE\textsuperscript{2} method \cite{feyel99} is theoretically capable of performing realistic two-scale simulations with arbitrary accuracy. Therefore, it serves as a reference method in the context of first-order homogenization based on the assumption of separated length scales. In the FE\textsuperscript{2}, the evaluations of the unknown macroscopic constitutive law are approximated by microscopic FE simulations. However, this comes along with computational costs that quickly exceed the capabilities of common workstations, both at present and in the foreseeable future. Roughly speaking, the computational effort required on the microscale multiplies with that of the macroscale, hence the method's name. It is thus worthwhile to develop order reduction methods for the microscopic problem.

A common approach within the field of computational homogenization (and well beyond) is to extract essential information from provided in silico data.
To this end, schemes based on the Proper Orthogonal Decomposition (POD) compute correlations within snapshot  data, \citep{sirovich1987}. \color{black} Such methods include the R3M~\cite{yvonnet2007} and can be further enhanced by the use of, e.g., the EIM, as in~\cite{radermacher2016}. Numerical comparisons of various schemes were conducted in \cite{Radermacher2013,Soldner2017}. To the best of the authors' knowledge, all published POD-based methods addressing the finite strain hyperelastic problem choose to reduce the number of degrees of freedom (DOF) of the displacement field. This results in sometimes significant speed-ups. Another important feature is that they allow for reconstruction of the microscopic displacement fields. The application of the snapshot POD to \emph{gradients} of the primal variables has been studied---e.g., for infinitesimal strain hyperelasticity \cite{Fritzen2018} and fluid mechanics~\cite{akkari2019}---but does not appear to have been investigated for finite strain hyperelasticity yet.\color{black}

Still, the solution of the reduced equations remains a complex task. It requires evaluations of material laws and numerical integration over the microstructure. Promising progress has been made in the field of efficient integration schemes, see for instance~\cite{An2009,hernandez2017}. A main reason for the speed-up of these methods is the reduced number of function evaluations.

The highest speed-ups are achievable if the computational effort of the determination of effective microstructural responses can be fully decoupled from underlying microstructural discretizations.
Such homogenization methods directly approximate the effective material law by means of a dedicated numerical scheme. Technically, this can be seen as the direct surrogation of unknown functions, e.g., of the effective free energy or stress. For instance, the Material Map~\cite{temizer2007} interpolates the coefficients of an assumed macroscopic material model. Another example is the NEXP method~\cite{yvonnet2013},  where \color{black} the effective stored energy density is approximated using a tensor product of one-dimensional splines. The authors treated the case of small strains by introducing the RNEXP method~\cite{Fritzen2018}, where the effective stored energy is interpolated by a dedicated kernel scheme.

However, interpolatory and regressional methods suffer the inherent drawback of not providing any explicit information on the microscale. For instance, microscopic displacement or stress fields cannot be reconstructed from the solutions of macroscopic interpolation. Another important open question is how to provide the supporting data points for the interpolation in an efficient manner. The data at these points is usually provided by the solution of a full-order model (FOM) and come along with the corresponding numerical costs. Hence, the positions of data points in the parameter space should be chosen carefully, as unnecessary or redundant solutions of the FOM should be avoided.
On the other hand, too sparsely seeded points might not capture the homogenized properties of the microstructure appropriately.

\subsection{Main Contributions and Outline}
\label{sec:intro:contributions}
The present work generalizes parts of the previous paper \color{black} \cite{Fritzen2018} to the finite strain regime. It aims at reducing the computational complexity  for the determination of the homogenized microstructural response, which is parametrized by the macroscopic deformation gradient acting as a boundary condition. This is achieved \color{black} by means of a \emph{Reduced Basis approximation of the microscopic deformation gradient}. The basis is obtained with the aid of a \emph{POD of snapshots of fluctuation fields of the deformation gradient}. Thus, the application of the RB model \emph{does not necessitate the computation of gradients of displacement fields}, and even does not require the displacements to be available at all. In other words, microscopic displacement fields are completely avoided. However, they \emph{can be reconstructed} from the RB approximation of the deformation gradient, uniquely up to rigid body motion.

Another key advantage is the \emph{sleek implementation} of the method. A  \emph{demonstration} \color{black} containing a minimum working example of the RB model with 50 lines of MATLAB/Octave code  is provided\color{black},~\citep{GitHubReducedBasisDemonstrator}.

As for the setup phase, the snapshot data is created by means of an \emph{efficient sampling procedure}  for the microscopic boundary condition\color{black}.
To this end,  the set of macroscopic \emph{Hencky strains} is identified as a suitable \emph{linear parameter space}\color{black}, within which the sampling sites are placed based upon \emph{physical interpretation}. This allows for \emph{control}
of the resolution of certain key characteristics of the effective material response while keeping the total number of samples within bounds.

The Reduced Basis method is presented in Section~\ref{sec:RB}. The basis identification is based on the sampling strategy developed in Section~\ref{sec:sampling}. Numerical examples are presented in Section~\ref{sec:numex}. Both the numerical and the theoretical findings are summarized and discussed in Section~\ref{sec:discussion}.

\subsection{Notation}
\label{sec:intro:notation}
The set of real numbers and the subset of positive numbers greater than zero are denoted by $\ffR$ and $\ffR_+$, respectively. Matrices are marked by two underlines and vectors by one underline, e.g., $\ull{A}$, $\underline{a}$. Vectors are assumed to be columns, and the dot product of two vectors of the same size is understood as the Euclidean scalar product, $\underline{x}\cdot\underline{y}=\underline{x}\TT\underline{y}$.
First order and second order tensors in coordinate-free description are denoted by bold letters, e.g., $\fA$, $\fa$. No conclusion can be drawn on the order of a tensor based on its capitalization. Here, the underlying space is always the Euclidean space $\ffR^3$ with its standard basis. First order and second order tensors can be represented as vectors and matrices, e.g., $\fA\leftrightarrow\underline{A}\in\ffR^3$ and $\fB\leftrightarrow\ull{B}\in\ffR^{3\times3}$, respectively. Norms of vectors and matrices respectively denote the Euclidean and the Frobenius norm. The norm of a tensor of second order equals the norm of its matrix representation for the chosen basis. Fourth order tensors are denoted by blackboard bold symbols other than $\ffR$, e.g., $\ffC$ and $\ffI$.
Components of tensors of order $M$ are with respect to the Euclidean tensorial basis $\fe^{(1)}\otimes\fe^{(2)}\otimes\dots\otimes\fe^{(M)}$, e.g., $A_{ij}$, $B_{ij}$ for second order tensors $\fA$, $\fB$ and $C_{ijkl}$,$C^\prime_{ijkl}$ for $\ffC$, $\ffC^\prime$. The following contractions are defined:
 \begin{align*}
 \fA\cdot\fB &= \sum_{i,j=1}^3 A_{ij}B_{ij}\,, &
    \ffC\cdot\fB &= \sum_{i,j,k,l=1}^3 C_{ijkl}B_{kl} \, \fe\bi\otimes\fe\bj\,,\\
 \fA\cdot\ffC &= \sum_{i,j,k,l=1}^3 A_{ij}C_{ijkl} \, \fe\bk\otimes\fe^{(l)}\,, &
    \ffC\cdot\ffC^\prime &= \sum_{i,j,k,l,m,n=1}^3 C_{ijmn}C_{mnkl}^\prime \, \fe\bi\otimes\fe\bj\otimes\fe\bk\otimes\fe^{(l)}\,.
\end{align*}\color{black}

\begin{figure}[H]
 \centering\includegraphics[scale=1]{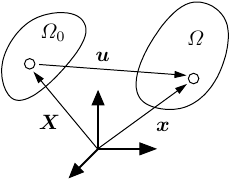}
 \caption{Initial ($\varOmega_0$) and current ($\varOmega$) configuration, with elementary kinematic quantities.}
 \label{fig:deformation}
\end{figure}

Let $\varOmega\subset\ffR^3$ be the domain occupied by a physical body undergoing elastic deformations, and let $\varOmega_0$ be its initial configuration. Then, $\fx$ and $\fX$ describe the coordinates of material points within the current configuration $\varOmega$ and within the reference state $\varOmega_0$, respectively. Their difference is the displacement $\fu=\fx-\fX$, see Figure~\ref{fig:deformation}.
The gradient of a vector field $\fv=\fv(\fX)$ is defined as a \emph{right gradient} and denoted by $\pd{\fv}{\fX}=\fv\otimes\nabla_{\rm X}$. The divergence of a second order tensor field is the vector field resulting from \emph{row-wise} divergence.
The boundaries of the respective configurations are denoted by $\partial\varOmega$ and $\partial\varOmega_0$. The set of square-integrable Lebesgue functions on the reference domain is tagged $L^2(\varOmega_0)$.

The displacement gradient $\fH=\fu\otimes\nabla_{\rm X}$ and the deformation gradient $\fF=\fx\otimes\nabla_{\rm X}$ are related through $\fF = \fH + \fI$, where $\fI$ is the second order identity tensor in three dimensions.
The determinant $J=\det{\fF}$ measures the relative volumetric change due to the present deformation.

Unimodular quantities, i.e., second order tensors with determinant ones, may be emphasized by a hat, e.g., $\widehat{\fF}=J^{-1/3}\fF$. This multiplicative decomposition is sometimes attributed to Flory~\cite{flory1961} and also goes by the name Dilatational-Deviatoric 
Multiplicative Split (DDMS). 

In the two-scale context, overlined symbols represent quantities on the macroscopic scale, e.g., $\ol\fA$, $\ol\fa$, while symbols without overline correspond to their microscopic counterpart, e.g., $\fA$, $\fa$. Equivalently, macroscopic quantities are called \emph{global} and microscopic ones are called \emph{local}. The volume average of a general local field $\varphi$
\begin{align}\label{eq:intro:averaging}
 \left<\varphi\right> &=\left< \varphi(\bullet) \right> = \frac{1}{\vert\varOmega_0\vert}\int_{\varOmega_0} \varphi(\bullet) \,{\rm d} V
\end{align}
is essential to the theory. The dependence of a microscopic quantity $\fA$ on both the microscopic coordinates $\fX$ and a macroscopic quantity $\ol\fB$ is denoted by $\fA=\fA(\fX;\ol\fB)$. In such a case, the components of the macroscopic quantity $\ol\fB$ are called \emph{parameters} of the microscopic function $\fA(\bullet;\ol\fB)$. The application of the volume averaging operator is abbreviated by  $\left<\fA\right>=\left<\fA(\bullet;\ol\fB)\right>$. The case of a concatenated function $f(\fA)=f(\fA(\fX;\ol\fB))$ is analogous, i.e. $\left<f\right>=\left<f(\fA)\right>=\left<f(\fA(\bullet;\ol\fB))\right>$, regardless of the tensorial order of the image of the function $f$.

\subsection{Material Models}
\label{sec:intro:materialmodels}
In this work, hyperelastic materials are investigated. They are characterized by \emph{stored energy density functions}~$W=W(\fF)$. The first Piola--Kirchhoff stress
\begin{align}\label{eq:intro:P}
 \fP(\fF)&=\pd{W}{\fF}(\fF)
\intertext{and the corresponding fourth-order stiffness tensor}
\label{eq:intro:ffC}
 \ffC(\fF)&=\pdII{W}{\fF}(\fF)
\end{align}
characterize the material response.

Henceforth, for reasons of readability, the stored energy density function $W$ will be spoken of as an energy, and the terms \emph{stored} and \emph{density} will not always be mentioned explicitly.
In the infinitesimal strain framework, hyperelastic energies have been formulated to model \color{black} deformation plasticity \citep[e.g.,][]{Fritzen2018,yvonnet2013,Bilger2005}. Although these models are only valid for purely proportional loading conditions, they provide means to simulate highly nonlinear material behavior in certain scenarios comparably easily within the context of hyperelasticity. Note that genuine dissipative processes require additional state describing variables with corresponding evolution laws.

The proposed method is \emph{suitable for any type of hyperelastic constitutive law}. As the modeling of complex material behavior is not the main focus of this study, the Neo--Hookean law
\begin{align}\label{eq:neohooke}
 W(\fF) = W_{\rm DDMS}(J,\widehat{\fF}) &= \frac{K}{4}\left[(J-1)^2+({\rm ln}\,J)^2\right] + \frac{G}{2} \lb\tr{\,\widehat{\fF}\TT\widehat{\fF}}-3\rb
\end{align}
is used, with $K$ the bulk modulus and $G$ the shear modulus. The volumetric part of the energy is taken from \cite{Doll1999}. Using the DDMS, a decoupled dependence on the volumetric and isochoric part of the deformation is assumed, which is a common way to model the distinct material behavior with respect to these two contributions, see e.g.,~\citep{Simo1988a}.

\subsection{Problem Setting of First Order Homogenization}
\label{sec:intro:problemsetting}
Assuming stationarity and separability of scales, the following coupled and deformation-driven problems can be derived by means of asymptotic expansion of the displacement $\fu$ and subsequent first order approximation. This procedure is carried out in \cite{Pruchnicki1998} with much detail. Here, the technical process is omitted and only the resulting equations are stated.

\subsubsection{Macroscopic Problem}
Balance of linear momentum
\begin{align}\label{eq:intro:macro:problem}
 {\rm Div}_{\rm \ol{X}}(\olfP) +\ol\fb &= \fzero,
\end{align}
where $\ol\fb$ denote bulk forces, and balance of angular momentum
\begin{align}\label{eq:intro:macro:problem_angular}
 \olfF^{-1}\olfP &= \olfP\TT\olfF^{\sf -T}\text{,}
\end{align}
along with well-posed boundary conditions that constitute the \emph{macroscopic boundary value problem}. This system of equations is closed by means of the hyperelasticity law, cf. \eqref{eq:intro:P},
\begin{align}\label{eq:intro:macro:P}
 \olfP(\olfF) &= \pd{\olW}{\olfF}.
\end{align}
Note that, in general, $\olW$ is a priori not available and $\eqref{eq:intro:macro:P}$ is thus a purely formal relation. For reasons of readability, the dependence of any quantity on the macroscopic material coordinate $\ol\fX$ is usually spared, e.g., $\olfF=\olfF(\ol\fX)$, $\olfH=\olfH(\ol\fX)$, or $\ol\fu=\ol\fu(\ol\fX)$.

\subsubsection{Microscopic Problem}
The \emph{microscopic boundary value problem} is given by the balance equations
\begin{align}\label{eq:intro:micro:problem}
 {\rm Div}_{\rm X}(\fP) &= \fzero\\ \label{eq:intro:micro:problem_angular}
 \fF^{-1}\fP &= \fP\TT\fF^{\sf -T}
\end{align}
and suitable boundary conditions, e.g., as discussed in \cite{miehe2003b}. In this work, \emph{periodic displacement fluctuation boundary conditions} are employed. The microscopic displacements take the form
\begin{align}\label{eq:intro:micro:displacement}
 \fu(\fX;\olfF) &= \ol\fu + \ol\fH\fX + \fw(\fX;\olfF).
\end{align}
Therein, the macroscopic displacement is independent of microscopic quantities. The second term, $\olfH\fX$, corresponds to a homogeneous deformation of the microstructure. The third term, $\fw(\fX)$, is a \emph{displacement fluctuation} with the \emph{zero mean property} $\langle\fw\rangle=\fzero$. Hence, the deformation gradient reads
\begin{align}\label{eq:intro:micro:defograd}
 \fF(\fX;\olfF) &= \olfF + \widetilde{\fH}(\fX;\olfF) = \olfF + \widetilde{\fF}(\fX;\olfF),
\end{align}
where the  fluctuation \color{black}part $\widetilde{\fH}=\fw\otimes\nabla_{\rm X}=\widetilde{\fF}$, has the zero mean property
\begin{align}\label{eq:intro:micro:fluct}
 \left<\widetilde{\fF}\right> &= \fzero .
\end{align}
Thus, the volume average of the local deformation gradient equals its macroscopic counterpart,
\begin{align}\label{eq:intro:micro:boundarycondition}
 \olfF &= \left<\fF\right>.
\end{align}
This motivates the identification of $\olfF$ as \emph{the boundary condition} to the microscopic problem~\eqref{eq:intro:micro:problem}.  
As for the material response, the following relationships can be deduced:
\begin{align}\label{eq:intro:micro:coupling_energy}
 \olW&=\left<W\right>,\\
 \label{eq:intro:micro:coupling_stress}
 \olfP &= \left<\fP\right>,\\
 \label{eq:intro:micro:coupling_stiffness}
 \olffC&\neq\left<\ffC\right>.
\end{align}
Equations~\eqref{eq:intro:micro:boundarycondition} and \eqref{eq:intro:micro:coupling_stress} are called kinematic and static \emph{coupling relations}, respectively. The inequality \eqref{eq:intro:micro:coupling_stiffness} generally holds for heterogeneous microstructures, even in the most simple case of infinitesimal strains and linear elasticity. More precisely, the volume average overestimates the effective stiffness in the spectral sense,
\begin{align}\label{eq:intro:micro:estimate_stiffness}
 \fx\cdot\left<\ffC\right>\cdot\fx &\geq \fx\cdot\olffC\cdot\fx \qquad \forall\text{ second order tensors }\fx.
\end{align}

\section{Reduced Basis Homogenization for Hyperelasticity}
\label{sec:RB}

\subsection{Formulation}
\label{sec:RB:formulation}
The \emph{Reduced Basis (RB)} scheme is based on a direct approximation of the microscopic deformation gradient~$\fF$ from Equation~\eqref{eq:intro:micro:defograd} without the need to explicitly have the corresponding displacement at hand.
The initial approximation is given by
\begin{align} \label{eq:RB:ansatz_with_xi}
 \fF_\xi(\fX;\olfF,\underline{\xi}) &= \olfF + \sum_{i=1}^N \fB\bi(\fX) \xi_i.
\end{align}
The  arguments \color{black} $\olfF$ and $\underline{\xi}\in\ffR^N$ are the boundary condition to \eqref{eq:intro:micro:problem} and the reduced coefficient vector, respectively. Note that the macroscopic coordinate $\ol\fX$ is not assumed to  influence the RB approximation\color{black}, i.e., the same approximation is made throughout the macrostructure. The set $\{\fB\bi\}_{i=1}^N$ is linearly independent within the space $L^2(\varOmega_0)$ and is called \emph{RB of $\fF$}. In a later context, it will also be referred to as the set of \emph{ansatz functions}.
In order to enforce the relationship
\begin{align} \label{eq:RB:upscaling}
 \left<\fF_\xi\right> &= \olfF
\end{align}
regardless of the reduced coefficients~$\underline{\xi}$, the basis functions are asserted to have the \emph{fluctuation} property, i.e., for $i=1,\dots,N$
\begin{align} \label{eq:RB:fluctuation}
 \left<\fB\bi\right> &= \fzero.
\end{align}
For now, the RB is assumed to be given.

The ansatz \eqref{eq:RB:ansatz_with_xi} allows for evaluation of the energy at the macroscale as a function of the macroscopic kinematic variable~$\olfF$ and of the reduced coefficients~$\underline{\xi}$, 
\begin{align} \label{eq:RB:energy1}
 \olW_\xi(\olfF,\underline{\xi}) &= \left< W(\fF_\xi) \right>.
\end{align}
By the principle of minimum energy, the optimal coefficients
\begin{align} \label{eq:RB:coefficients}
 \underline{\xi}^*(\olfF) &= \arg\min_{\underline{\xi}\in\ffR^N} \olW_\xi(\olfF,\underline{\xi})
\end{align}
are sought after. The unconstrained and unique solvability of this task is assumed for the moment and will be addressed in Section~\ref{sec:RB:details}.
The solution of \eqref{eq:RB:coefficients} defines the \emph{RB approximation} of the deformation gradient
\begin{align} \label{eq:RB:ansatz}
 \fF\RB(\fX;\olfF) &= \olfF + \sum_{i=1}^N \fB\bi(\fX) \xi_i^*(\olfF).
\end{align}

The microscopic energy, stress, and stiffness within the microstructure are then approximated by
\begin{align}
 \label{eq:RB:WRB} W\RB(\fX;\olfF) &= W(\fF\RB(\fX;\olfF)),\\
 \label{eq:RB:PRB} \fP\RB(\fX;\olfF) &= \pd{W}{\fF}(\fF\RB(\fX;\olfF)),\\
 \label{eq:RB:CRB} \ffC\RB(\fX;\olfF) &= \pdII{W}{\fF}(\fF\RB(\fX;\olfF)),
\end{align}
respectively. The resulting effective energy is readily given by
\begin{align}
 \label{eq:RB:energy}
 \olW\RB(\olfF) &= \left< W\RB \right>.
\end{align}
Also, the effective responses $\olfP\RB(\olfF)$ and $\olffC\RB(\olfF)$ may now be calculated. However, before going into detail on that, it is advantageous to first elaborate on the solution of the minimization problem \eqref{eq:RB:coefficients}. This short survey will reveal essential properties of some occurring quantities that are important for the determination of the effective material response.

The necessary, first order optimality conditions to \eqref{eq:RB:coefficients} define the components of the residual vector $\underline{r}\in\ffR^N$,
\begin{align} \label{eq:RB:residual}
 r_i(\olfF,\underline{\xi}) &= \pd{\olW_\xi}{\xi_i}(\olfF,\underline{\xi}) = \left< \pd{W}{\fF}(\fF_\xi)\cdot\pd{\fF_\xi}{\xi_i} \right> = \left< \fP(\fF_\xi)\cdot\fB\bi \right> = 0 \qquad (i=1,\dots,N).
\end{align}
\begin{note}\label{note:orthogonality}
 The solution stress field $\fP\RB$ 
 is $L^2(\varOmega_0)$-orthogonal to the RB ansatz functions $\{\fB\bi\}_{i=1}^N$.
\end{note}

A viable choice for the solution of the minimization problem \eqref{eq:RB:coefficients} is the Newton--Raphson scheme, which necessitates the Jacobian $\ull{D}\in{\rm Sym}(\ffR^{N\times N})$ with the components
\begin{align}\label{eq:RB:Jacobian}
 D_{ij} &= \left< \fB\bi \cdot \pd{\fP}{\xi_j}(\fF_\xi) \right> = \left< \fB\bi \cdot \ffC(\fF_\xi) \cdot \fB\bj \right>=D_{ji} \qquad(i,j=1,\dots,N)\,.
\end{align}
Then, the $k$\textsuperscript{th} iteration to the solution $\underline{\xi}^*(\olfF)$ reads
\begin{align}
 \underline{\xi}\bk &= \underline{\xi}^{(k-1)} - \lb\ull{D}^{(k-1)}\rb^{-1}\underline{r}^{(k-1)} \qquad (k\geq1)\,.
\end{align}
The initial guess $\underline{\xi}^{(0)}$ can be zero or a more sophisticated alternative.

The deduction of the effective material response by means of Note~\ref{note:orthogonality} and the Jacobian $\ull{D}$ is given in Appendix~\ref{app:effective}. The following list summarizes the homogenized quantities arising from the $\fF$-RB:
\begin{align*} 
 \olfF &= \left<\fF\RB\right>&&\text{see }\eqref{eq:RB:upscaling}\\
 \olW\RB &= \left< W\RB \right>&&\text{see }\eqref{eq:RB:energy}\\
 \olfP\RB &= \left< \fP\RB \right> &&\text{see Appendix~\ref{app:effective:stress}}\\
 \olffC\RB &= \left< \ffC\RB \right> - \sum_{i,j=1}^N D_{ij}^{-1} \left< \ffC\RB \cdot \fB\bi \right> \otimes \left< \fB\bj \cdot \ffC\RB \right> &&\text{see Appendix~\ref{app:effective:stiffness}}
\end{align*}

In total, the problem of determining both the local field $\fF$ and the homogenized material properties depends only on $N$ degrees of freedom, namely on the $N$ coefficients $\xi_i$. Usually, the number of DOF $N$ is in the range of 10 to 150, which compares to the full order model's complexity that can easily reach more than 10\textsuperscript{5} DOF.

\begin{Remark}Despite this impressive reduction of the number of DOF, the computational costs associated with the assembly of the residual $\underline{r}$ and of the Jacobian $\ull{D}$ still relate to the number of quadrature points of the microstructural discretization.\end{Remark}

This method extends corresponding methods for the geometrically linear case, where the infinitesimal strain tensor $\feps={\rm sym}(\fH)$ is considered. For more information on these topics, the reader is referred to the authors' previous work~\cite{Fritzen2018} or standard literature, such as~\citep[part II.C]{Castaneda1998}.

\subsection{Identification of the Reduced Basis}
\label{sec:RB:identification}
The basis $\{\fB\bi\}_{i=1}^N$ is computed by means of a classical snapshot  POD. \color{black} In contrast to many other POD based reduction methods, it is important to point out that here, the primal variable is \emph{not} taken to be the displacement field , $\fu$\color{black}. Instead, the POD is performed on  deformation gradient fluctuations, $\widetilde{\fF}$.\color{black}

During the snapshot POD, \emph{snapshots} are first created by means of high-fidelity solutions to the nonlinear microscopic problem~\eqref{eq:intro:micro:problem} with different \emph{snapshot boundary conditions} $\olfF\bj$, $j=1,\dots,N_{\rm s}$, which are also called \emph{training boundary conditions}. Each of these boundary conditions leads to a solution field $\fF\bj(\fX;\olfF)$. Typically, the Finite Element Method (FEM) or solvers making use of the Fast Fourier Transform \citep[e.g.,][]{kabel2014} are used to this end. The RB method presented here is independent of the discretization method utilized to obtain full field solutions. It is merely necessary to know the  quadrature weights and the related discrete values of the solutions $\fF\bj(\fX;\olfF\bj)$. For better readability, the continuous notation is maintained for the moment. The corresponding fluctuation fields are computed by means of local subtraction of the macroscopic deformation gradient
\begin{align}\label{eq:RB:identification:fluctuations}
 \widetilde{\fF}\bj(\fX;\olfF\bj) &= \fF\bj(\fX;\olfF\bj) - \olfF\bj \qquad (j=1,\dots,N_{\rm s}).
\end{align}
Each of these $N_{\rm s}$ fluctuation fields $\widetilde{\fF}\bj$ represent one snapshot.

Second, the most dominant features of the snapshots are extracted. This is done by means of the eigendecomposition of the correlation matrix. It consists of the mutual $L^2(\varOmega_0)$ scalar products of the snapshots, $\big< \widetilde{\fF}\bi \cdot \widetilde{\fF}\bj \big>$ ($i,j=1,\dots,N_{\rm s}$), cf. \eqref{eq:intro:averaging}. The remaining procedure is standard, see for instance \citep{sirovich1987} or \citep{quarteroni:2016}: the $N_{\rm s}$ eigenvalues $\lambda_j$, corresponding to the eigenvectors $\underline{E}\bj\in\ffR^{N_{\rm s}}$, are sorted in descending order and truncated by only considering the first $N$ values, $\lambda_1\geq\ldots\geq\lambda_N$. The decision on a particular threshold index $N$ is based on consideration of the Schmidt--Eckhard--Young--Mirsky theorem. Finally, the RB is constructed as
\begin{align}\label{eq:RB:identification:definiton}
 \fB\bi(\fX) &= \sum_{j=1}^{N_{\rm s}} \frac{1}{\sqrt{\lambda_i}}\lb\underline{E}\bi\rb_j \widetilde{\fF}\bj(\fX) \quad (i=1,\ldots,N)
\end{align}
where the factor $1/\sqrt{\lambda_i}$ accounts for $L^2(\varOmega_0)$ normalization of the base elements.
We conclude that the RB is a collection of $L^2(\varOmega_0)$ orthonormal $\widetilde{\fH}$-like quantities.

\subsection{Mathematical Motivation of the Reduced Basis model}
\label{sec:RB:justification}
Next, the obtained deformation gradient field $\fF\RB$ and the associated stress field $\fP\RB$ are shown to weakly solve the original problem \eqref{eq:intro:micro:problem}, ${\rm Div}_{\rm X}(\fP)=\fzero$.

Let $\delta\fw\in H^1_0(\varOmega_0)$ be an admissible test function, i.e., a once weakly differentiable periodic displacement fluctuation field, and let $\delta\fH=\delta\fw\otimes\nabla_{\rm X}$ denote its gradient. Then, the well-known weak form of~\eqref{eq:intro:micro:problem} is equivalent to the principle of virtual work,
\begin{align}\label{eq:RB:weakform}
 \delta \olW = & \int_{\varOmega_0} \fP\cdot\delta\fH\,{\rm d}V = 0.
\end{align}
 The residual~$\underline{r}$ from \eqref{eq:RB:residual} coincides with the integral of the weak form, if the test function $\delta\fw$ is chosen suitably. 
As the basis elements $\fB\bi$ are linear combinations of deformation gradient fluctuations $\fw\bj\otimes\nabla_{\rm X}$, cf. \eqref{eq:RB:identification:definiton}, the equivalence of \eqref{eq:RB:residual} and \eqref{eq:RB:weakform} is obvious.\color{black}

It follows that the Reduced Basis scheme is a \emph{Galerkin} procedure, taking the displacement fields
corresponding to the RB elements $\fB\bi$ as both ansatz and test functions. 
Hence, the \emph{RB model is equivalent to the FEM}, but with basis functions that are globally supported in $\varOmega_0\backslash\partial\varOmega_0$.
In other words, the basis functions of the RB method  span a subspace \color{black} of dimension $N$ within the high-dimensional
space of FE basis functions. Although this property is shared with RB schemes based on POD of displacement snapshots, a notable difference is that this novel approach directly operates on fields entering the constitutive equations.

\subsection{Details on the Coefficient Optimization}
\label{sec:RB:details}
The coefficient optimization task \eqref{eq:RB:coefficients} leads to a weak solution of the microscopic boundary value problem, as was just shown. Hence, the well-established theories on which the FEM is built, e.g., the calculus of variations, are applicable to the presented method just as much. This implies that the well-known issues with suitable convexity conditions and with existence and uniqueness of minimizers apply to the RB method, too. We focus on ad hoc numerical treatments of these issues. For more details on the theoretical part, the reader is referred to standard literature, such as~\cite{Ball1976}, and recent developments in this matter, e.g.,~\cite{schneider2017}.

A constraint to the optimization problem is the physical condition
\begin{align}\label{eq:RB:Jpos}
 J &= {\rm det}(\fF_\xi(\fX))>0\quad\forall\,\fX\in\varOmega_0,
\end{align}
which means that no singular ($J=0$) or self-penetrating ($J<0$) deformations shall occur. This reduces the set of admissible coefficients $\underline\xi$ to a subset of $\ffR^N$ that is nontrivial to access.
The positiveness of the microscopic determinant of the deformation gradient introduces a constraint to the, thus far, unconstrained minimum problem \eqref{eq:RB:coefficients}, representing the weak form of \eqref{eq:intro:micro:problem} in the RB setting.

In case of a violation of the inequality~\eqref{eq:RB:Jpos}, the implementation of the RB method is prone to failure as soon as the constitutive law \eqref{eq:neohooke} is evaluated. This occurs only in the context of volume averaging, i.e., for the assembly of the residual, the Jacobian, or the effective energy, stress, or stiffness. The numerical quadrature approximating the volume averaging operation is
 \begin{align}\label{eq:RB:quadrature}
  \left<\bullet\right> &= \frac{1}{|\varOmega_0|}\int_{\varOmega_0} \bullet(\fX) \,{\rm d}V \approx \frac{1}{|\varOmega_0|}\sum_{p=1}^{N_{\rm qp}} \bullet(\fX_p)\,w_p.
 \end{align}
Here, $N_{\rm qp}$, $\fX_p$, and $w_p$ respectively denote the number of quadrature points, their positions, and the corresponding positive weights.
Moreover, even if the inequality \eqref{eq:RB:Jpos} is satisfied everywhere, the local field $\fF_\xi$ might possibly have some positive but overly small values of the determinant, $0<{\rm det}(\fF_\xi(\fX))\ll1$, that are unphysical. In such a case, the energy optimization, cf. \eqref{eq:RB:coefficients}, would be dominated by these nearly singular points. Instead of allowing the optimization to focus almost exclusively on these exceptional points, we interpret unphysically small values of the determinant as limitations to the reliability of the RB method. On the other hand, large values ${\rm det}(\fF_\xi(\fX))\gg1$ are not too detrimental to the functionality of the scheme, although such values are just as questionable.

\begin{figure}[H]
 \centering\includegraphics[scale=1]{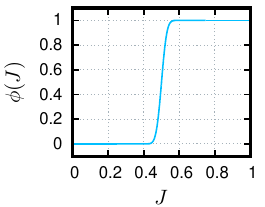}
 \caption{Cutoff function $\phi$. Its value is used as a reliability factor in the numerical quadrature.}
 \label{fig:RB:cutoff}
\end{figure}

Thus, the following \emph{weighted numerical quadrature rule} is introduced:
 \begin{align}\label{eq:RB:weighted_averaging}
  \left<\bullet\right> &\approx \lb\sum_{p=1}^{N_{\rm qp}} \bullet(\fX_p)\,\phi(J_p)\,w_p\rb/\lb\sum_{p=1}^{N_{\rm qp}}\phi(J_p)w_p\rb.
 \end{align}
Therein, the almost smooth \emph{cutoff function $\phi:\ffR\to\ffR_{\geq0}$} is empirically defined by
\begin{align}\label{eq:RB:cutoff}
 \phi(J) &=
    \begin{cases}
        1                                           & \text{if $J>0.6$}\\
        0.5\,{\rm erf}(30\,J-15)+0.5    & \text{if $0.4<J\leq0.6$}\\
        0                                           & \text{if $J\leq0.4$}
    \end{cases}\,\text{.}
\end{align}
which makes use of the well-known error function. The cutoff function is depicted in Figure~\ref{fig:RB:cutoff}. This reliability indicator could, in principle, be modified, e.g., the steepness, the smoothness, and its center could be adjusted. Thus, it should be regarded as an example only.

This weighted numerical quadrature rule is used henceforward for all \emph{numerical} volume averaging operations. Its application will not be noted explicitly. However, the \emph{theoretical} derivation of the RB method, as described in Section~\ref{sec:RB:formulation}, is not affected by this, i.e., volume averaging operations remain exact as far as the theory is concerned. The two most important consequences of this numerical tweak are:
\begin{itemize}
 \item The RB method is robust with respect to outlier values of the determinant. The modified quadrature rule extends the set of coefficient vectors $\underline{\xi}$ for which effective quantities can be computed, albeit approximately, to the whole space $\ffR^N$. 
 \item The significance of local fields varies with the value of the cutoff function. When $\phi$ attains values less than one, information is considered accordingly less reliable. In this sense, microscopic information is filtered based on a trust region for $J$ defined by $\phi$ can be seen as a reliability indicator.
\end{itemize}

\section{Sampling}
\label{sec:sampling}

\subsection{General Considerations}
\label{sec:sampling:general}
The proposed sampling strategy builds on the previous work~\cite{Fritzen2018}, in which the authors proposed an analogous sampling procedure for the small strain setting. However, substantial modifications are required in order to account for the finite rotational part, $\ol\fR$, of the macroscopic deformation gradient, $\olfF$, and the nonlinearity of the volumetric part of the deformation, $J$, with respect to the local displacements, $\fu$.
For the setup of the Reduced Basis model as described in Section~\ref{sec:RB:identification}, the space of macroscopic deformation gradients,
\begin{align}\label{eq:sampling:cFbar}
 \ol\cF &= \{\text{second order tensors }\ol\fF\,|\,\det{\ol\fF}>0 \},
\end{align}
needs to be sampled, i.e., the discrete sampling set $\ol\cF_{\rm s} = \{\olfF\bm\}_{m=1}^{N_{\rm s}}\subset\ol{\cF}$ is to be defined. Two contradictory requirements need to be satisfied when constructing $\ol\cF_{\rm s}$:
\begin{enumerate}[leftmargin=*,labelsep=4.9mm]
 \item The samples should be \emph{densely and homogeneously distributed} within the space of all admissible macroscopic kinematic configurations. This is owing to the desire that the POD may extract correlation information from a holistic and unbiased set. In other words, the samples should be as \emph{uniformly random} as possible \emph{within the anticipated query domain of the surrogate}.
 \item The sample number $N_{\rm s}$ should not exceed a certain limit. Only with this property may the RB be identified within the bounds of available computational resources (e.g., memory and CPU time).
\end{enumerate}

\subsection{Large Strain Sampling Strategy}
\label{sec:sampling:largestrain}
The set of admissible macroscopic deformation gradients $\ol\cF$ is a subset of a nine-dimensional space ($\ol{\ull F}\in\ffR^{3\times3}\sim\ffR^9$) restricted by the inequality 
\begin{align}\label{eq:sampling:determinant_condition}
 \det{\olfF}>0.
\end{align}
Therefore, a regular grid in the components of $\olfF$ might lead to a prohibitively large amount of samples, and even to a violation of~\eqref{eq:sampling:determinant_condition}. For instance, such a grid with a rather moderate resolution of just 10 samples of each component would require 1 billion solutions \color{black} of the FOM\color{black}. Also, the subsequent POD would involve a snapshot matrix and/or correlation matrix of accordingly huge dimensionality.

In order to decrease the dimension of the sampling space, recall the polar decomposition of the deformation gradient, $\olfF=\ol\fR\,\olfU$, where $\ol\fR$ is a rotation and $\olfU$ is the symmetric positive definite (s.p.d.) stretch tensor. Material objectivity implies the energy function to be independent of the frame of reference,
\begin{align}\label{eq:sampling:objectivity:W}
 \olW(\ol\fR\,\olfU) &= \olW(\olfU),
\intertext{and the transformation behavior}
\label{eq:sampling:objectivity:P}
  \olfP(\ol\fR\,\olfU) &= \ol\fR\,\olfP(\olfU)
\intertext{for the first Piola--Kirchhoff stress and}
\label{eq:sampling:objectivity:ffC}
    \ol C_{ijkl}(\ol\fR\,\olfU)&= \sum_{m,n=1}^3 \delta_{ik}\ol U^{-1}_{lm}\ol P_{mj}(\olfU) + \ol R_{im}\ol C_{mjnl}(\olfU)\ol R_{kn}\qquad(i,j,k,l=1,2,3)
\end{align}
for the components of the corresponding stiffness tensor $\ffC$, see Appendix~\ref{app:objectivity}.
 These known facts lead to\color{black}
\begin{note}\label{note:sampling:U}
 In order to collect representative samples of the hyperelastic response functions $\olW$, $\olfP$, and $\olffC$, it suffices to evaluate them on samples of the stretch tensor $\olfU\sim\ffR^6$ instead of evaluating them on samples of the deformation gradient $\olfF\sim\ffR^9$.
\end{note}
This effectively reduces the number of dimensions of $\ol\cF$ from nine to six.
The same dimensionality is attained when considering the response functions with respect to a symmetric measure of strain, e.g., as is done in \cite{miehe1996} where the tangent stiffness is efficiently computed using a perturbation technique. However, such measures are unsuitable for reduction by means of the proposed RB method.

The remaining six-dimensional space of s.p.d. tensors is not linear but a convex cone (which does not include the zero element). Moreover, linearly combining elements within this space quickly leads to values of $\ol J=\det{\olfU}$ describing unphysically large changes in volume.  For instance, $\olfU=1.3\,\fI$ equates to more than 100\% volumetric extension\color{black}, which is well beyond the regime of usual hyperelastic materials that are often nearly incompressible. On the other hand, 100\% deviatoric strain is within the range of many standard materials, such as rubber. Hence, \emph{in order to describe the space of practically relevant stretch tensors}, we propose to apply the DDMS to the macroscopic stretch tensor,
\begin{align}\label{eq:sampling:ddms}
 \olfU &= \ol{J}^{1/3}\,\widehat{\olfU}.
\end{align}
Let $\widehat{\ol\cU}$ denote the manifold of unimodular macroscopic stretch tensors $\widehat{\olfU}=(\ol{J})^{-1/3}\olfU$. The multiplicative split~\eqref{eq:sampling:ddms} is the basis for:
\begin{Proposition}\label{conc:sampling:det}
 The set of practically relevant macroscopic stretch tensors $\ol{\cU}$ can be sampled via sampling of both the macroscopic determinants,
 \begin{align*}
  \left\{\ol{J}\bm\right\}_{m=1}^{N_{\rm dil}} &\subset \ffR_+\text{ ,}
  \intertext{and the macroscopic unimodular stretch tensors,}
  \left\{\widehat{\olfU}\bj\right\}_{j=1}^{N_{\rm dev}} &\subset  \widehat{\ol\cU},
 \end{align*}
 where $N_{\rm dil}$ and $N_{\rm dev}$ are the numbers of the samples. \color{black}The sampling set is determined by the product set
 \begin{align}\label{eq:sampling:productset}
  \left\{\lb\ol{J}\bm\rb^{1/3}\widehat{\olfU}\bj\right\}_{m,j=1}^{m=N_{\rm dil},\,j=N_{\rm dev}}\subset\ol{\cU}.
 \end{align}
\end{Proposition}

The choice of the dilatational samples is relatively straightforward. For many common materials, the expected range of $\ol{J}$ is rather narrow, so a few equisized or adaptive sub-intervals around $\ol{J}=1$ deliver sufficient resolution.

For the space of unimodular s.p.d. matrices representing $\widehat{\olfU}\in\widehat{\ol{\cU}}$, basic results of Lie group theory can be exploited. We restrict to stating well-known facts that are necessary at this point. For more details, the interested reader is referred to standard text books, such as \cite{faraut2008}.
\begin{corollary}\label{cor:Lie}
 Let
 \begin{align*}
  \rm{symsl} &= \left\{ \ull{Y}\in\ffR^{3\times3}\,\big|\,\ull{Y}=\ull{Y}\TT, \tr{\ull{Y}}=0\right\}
  \intertext{be the \emph{tangent space} and}
  \rm{SymSL}_+ &= \Big\{ \ull{U}\in\ffR^{3\times3}\,\big|\,\ull{U}=\ull{U}\TT,\break \det{\ull{U}}=1, \underline{x}\TT\,\ull{U}\,\underline{x}>0 \,\forall\underline{x}\in\ffR^3 \Big\}
 \end{align*}
be the \emph{manifold} of unimodular s.p.d. matrices. The matrix exponential maps the tangent space bijectively onto the manifold,
\begin{align*}
 \rm{exp}: \quad\rm{symsl} \to \rm{SymSL}_+\,.
\end{align*}
\end{corollary}
The proof of this statement is standard, e.g., by means of the eigenvalue decomposition, and does not necessitate the reference to the abstract setting of Lie groups. In fact, all of the following arguments could be given without the notion of tangent spaces and manifolds. However, this notion is a fundamental concept in nonlinear mechanics. For a very descriptive and comprehensive work on this topic, the reader is referred to \cite{Neff2016}. We choose to build upon this concept, as it comes along with vivid interpretations of the function spaces $\ol\cU$  and $\widehat{\ol \cU}$.\color{black}

The set $\rm SymSL_+$ is the set of matrix representations of  the stretch tensors $\widehat{\olfU}\in\widehat{\ol{\cU}}$. The tangent space $\rm{symsl}$ is the set of \emph{Hencky strains}, which is linear\color{black}. Hence, by virtue of the matrix exponential, the sampling of the nonlinear manifold $\widehat{\ol{\cU}}$ can be reduced to the sampling of a linear space. Moreover, the restrictions of symmetry and zero trace render the tangent space five-dimensional. This property is, by definition, shared with the manifold $\rm SymSL_+$.

\begin{figure}[H]\centering
 \includegraphics[height=7cm]{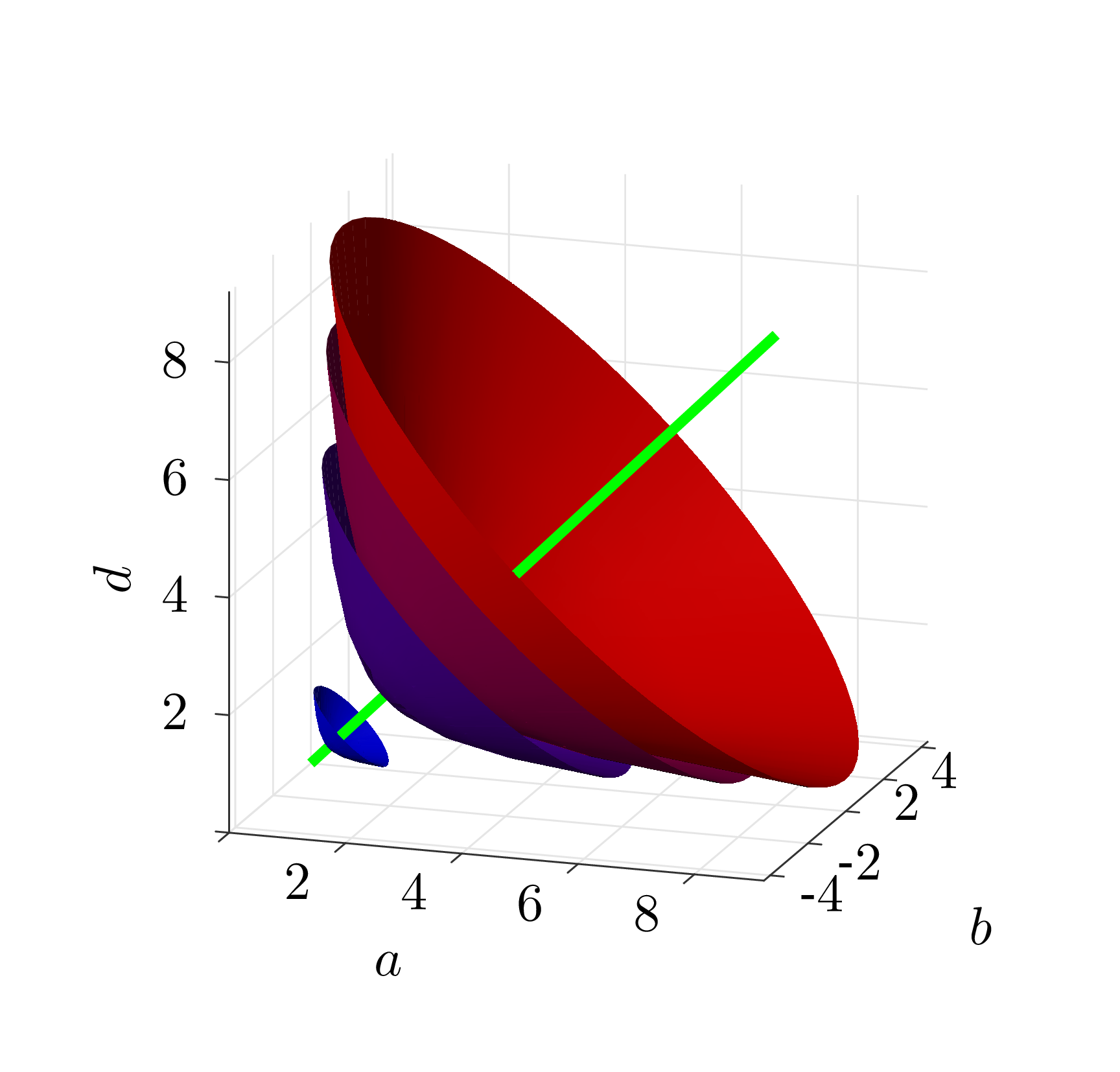}\includegraphics[height=7cm]{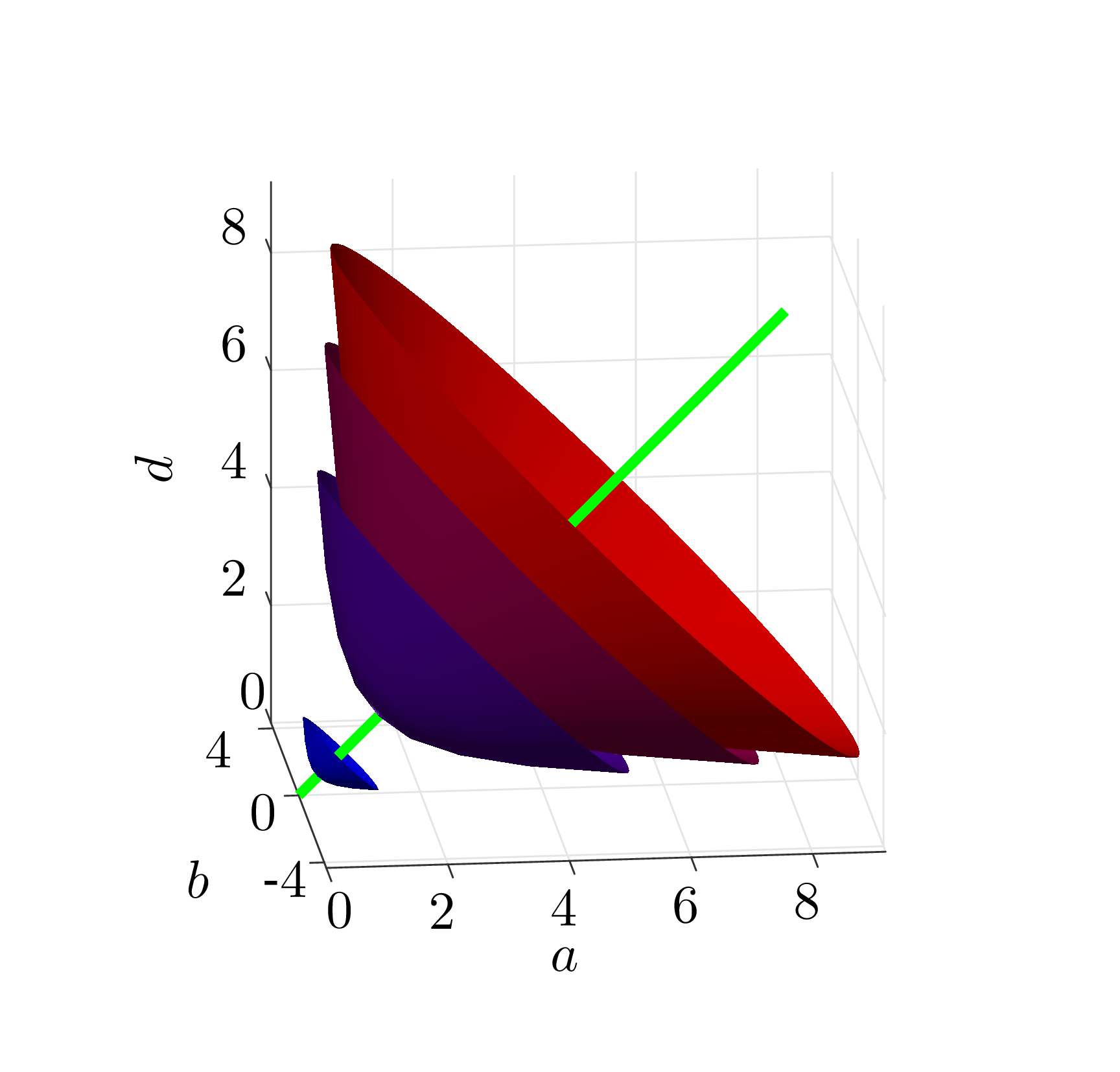}
 \caption{Visualizations of the family of $\olfU$-manifolds with constant determinant $\ol J\in\{0.1,\ldots,4.0\}$ for the two-dimensional case from two different perspectives. The green line represents the set $\{\lambda\fI\,|\,\lambda>0\}$.}
 \label{fig:sample_stretchtensor}
\end{figure}

The two-dimensional case is now addressed for the sake of visualization. In this setting, the nonlinearity of the manifold and the structure of the sampling set $\ol\cU$ can be illustrated figuratively. With the subscript (2) denoting two-dimensional quantities, a basis of the tangent space is given by
 \begin{align*}
  \ull{Y}^{(1)}_{(2)} &=\sqrt{\frac{1}{2}}\sV 1 & 0 \\ 0 & -1\eV \;\text{ and } & \ull{Y}^{(2)}_{(2)} &= \sqrt{\frac{1}{2}}\sV 0 & 1 \\ 1 & 0 \eV.
 \end{align*}
The stretch tensors are obtained through
 \begin{align*}
  \ull{\ol U}_{(2)} &= \ol{J}^{\frac{1}{2}}{\rm exp}\lb t\lb\alpha\ull{Y}^{(1)}_{(2)}+\beta\ull{Y}^{(2)}_{(2)}\rb\rb = \sV a&b\\b&d\eV.
 \end{align*}
A visualization of such samples is depicted in Figure~\ref{fig:sample_stretchtensor}. There, for the sake of visual clarity, the determinant $\ol J$ is sampled by four equidistant (and rather unrealistic) values between 0.1 and 4. The value $t\in[-2,2]$ is called \emph{deviatoric amplitude}. A densely uniform sampling $\varphi\in[0,2\pi)$ yields the coefficients $\alpha=\cos{\varphi}$ and $\beta=\sin{\varphi}$.

This emphasizes the important role of the DDMS in the context of sampling, as utilized in \eqref{eq:sampling:productset}---it allows for the identification of a physically meaningful sampling domain that is much smaller than the surrounding space of all admissible stretch tensors. On a side note, the isodet surfaces are perpendicular to the line representing the dilatational stretch tensors.

The proposed sampling procedure for $\ol{\cU}$ in three dimensions is given in Algorithm~\ref{algo:sampling}. For this purpose, an orthonormal basis $\ull{Y}^{(1)},\dots,\ull{Y}^{(5)}$ of the tangent space $\rm symsl$ is fixed, cf. Appendix~\ref{app:exp:basis}. The numbers of different kind of samples $N_{\rm det}$, $N_{\rm dir}$, and $N_{\rm amp}$ relate to the quantities $N_{\rm dil}$ and $N_{\rm dev}$ introduced in~\eqref{eq:sampling:productset} by $N_{\rm det}=N_{\rm dil}$ and $N_{\rm dir}N_{\rm amp}=N_{\rm dev}$.

\begin{algorithm}[H]
	\SetAlgoLined
	\SetEndCharOfAlgoLine{ }
	\Input{$\ol{J}_{\rm min},\ol{J}_{\rm max}$ minimum and maximum determinant with $\ol{J}_{\rm min}<1<\ol{J}_{\rm max}$\\$t_{\rm max}>0$ maximum deviatoric amplitude\\$N_{\rm det}$ number of macroscopic determinants\\$N_{\rm dir}$ number of deviatoric directions\\$N_{\rm amp}$ number of deviatoric amplitudes }
	\Output{$N_{\rm det}N_{\rm dir}N_{\rm amp}$ samples of $\olfU$}
	Place $N_{\rm det}$ determinants regularly between the extremal values,
	\begin{align*}
	 \ol{J}_{\rm min}=\ol{J}^{(1)}<\ldots<1<\ldots<\ol{J}^{\lb N_{\rm det}\rb}=\ol{J}_{\rm max}\,.
	\end{align*}\\
	Generate any approximately uniform distribution of $N_{\rm dir}$ directions in $\ffR^5$, e.g., as in \cite{Fritzen2018},
	\begin{align*}
	 \left\{\underline{N}\bn\right\}_{n=1}^{N_{\rm dir}}\subset\left\{\underline{N}\in\ffR^5\,:\, \lVert\underline{N}\rVert=1\right\}.
	\end{align*}\\
	Place $N_{\rm amp}$ deviatoric amplitudes regularly between 0 and the expected maximum value,
	\begin{align*}
	 0\leq t^{(1)}<\ldots<t^{\lb N_{\rm amp}\rb}=t_{\rm max}\,.
	\end{align*}\\
	Return the set of samples of $\olfU$:
	\begin{align}\label{eq:sampling:U}
	 \left\{ \lb\ol{J}\bm\rb^{1/3}{\rm exp}\lb t\bp \left[ \sum_{k=1}^5 \lb\underline{N}\bn\rb_k \ull{Y}^{(k)}  \right] \rb \right\}_{m,n,p=1}^{m=N_{\rm det},\,n=N_{\rm dir},\,p=N_{\rm amp}} \subset\ol{\cU}
	\end{align}
	\caption{Sampling of the macroscopic stretch tensor.}
	\label{algo:sampling}
\end{algorithm}

The order of Steps~1 to~3 is interchangeable. Details on these parts are now presented:
\begin{enumerate}[leftmargin=1.5cm,labelsep=4.9mm]
 \item[Step 1.] Uniform seeding of the determinants is actually not required, but any pattern implying the sampling determinants $\{\ol J\bk\}_{k=1}^{N_{\rm det}}$ to be dense in $[\ol J_{\rm min},\ol J_{\rm max}]$ as $N_{\rm det}\to\infty$ works without loss of generality. In this way, the dilatational response may be resolved adaptively.
 \item[Step 2.] The generation of uniform point distributions on spheres is a research topic on its own, see~\citep{Brauchart2015} for an overview. The method described in~\cite{Fritzen2018} is based on energy minimization, which is also used in the present work. Some point sets of various sizes are included in the example program~\citep{GitHubReducedBasisDemonstrator}. More detailed investigations on this topic and an open-source code of a point generation program are part of another work, \citep{kunc2018}. Alternatively, Equal Area Points~\citep{leopardi2006} may be used as a rough but quickly computable approximation of such point sets.
 \item[Step 3.] As in Step~1, the uniform placement of the deviatoric amplitudes, $t\bp$, may be substituted by adaptive alternatives. In~\citep{Fritzen2018}, we have suggested to use an exponential distance function.
\end{enumerate}

\begin{figure}[H]\centering
 \includegraphics[scale=1]{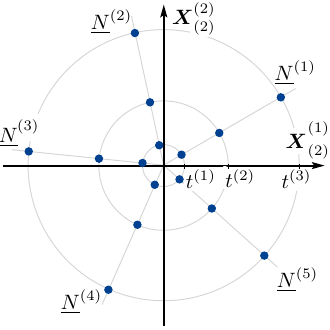}
 \caption{Example of a discretization of the two-dimensional tangent space. The samples are placed along the equidistant (in higher dimensions---uniformly distributed) directions with nonuniformly increasing amplitude.}
 \label{fig:sample_tangentspace}
\end{figure}

The result of Steps~2 and~3, i.e., the sampling of the tangent space, is exemplified in Figure~\ref{fig:sample_tangentspace} for the two-dimensional case ($\fu\in\ffR^2$) and for $N_{\rm dir}=5$ and $N_{\rm amp}=3$. There, an adaptive spacing of the deviatoric amplitudes is applied. This might be beneficial for capturing strongly changing material behavior near the relaxed state.

In general, the vector $\underline{N}\in\ffR^5$, $\lVert\underline{N}\rVert=1$, corresponding to a macroscopic Hencky strain \color{black} characterizes the \emph{direction of the applied stretch $\olfU$}, which we also coin the \emph{load case}.

 \subsection{Application of the Stretch Tensor Trained Reduced Basis Model}
 \label{sec:sampling:application}
 Since the RB model is trained on only the rotationally invariant part of $\olfF$ but should be applied to general deformation gradients, the transformation rules \eqref{eq:sampling:objectivity:P} and \eqref{eq:sampling:objectivity:ffC} are incorporated during the evaluation of the surrogate model. Details on the procedure are given in Algorithm~\ref{algo:F}.
\begin{algorithm}[H]
	\SetAlgoLined
	\SetEndCharOfAlgoLine{ }
	\Input{$\olfF$ macroscopic deformation gradient}
	\Output{$\olfP\RB(\olfF)$, $\olffC\RB(\olfF)$ effective material response}
	Compute polar decomposition $\olfF=\ol\fR\,\olfU$.\\
	Compute approximations of effective stress $\olfP\RB(\olfU)$ and effective stiffness $\olffC\RB(\olfU)$.\\
	Transform effective responses to correct frame $\olfP\RB(\olfF)$, $\olffC\RB(\olfF)$, using $\ol\fR$, cf. \eqref{eq:sampling:objectivity:P}, \eqref{eq:sampling:objectivity:ffC}.
	\caption{Online phase of the stretch tensor trained Reduced Basis method}
	\label{algo:F}
\end{algorithm}

\section{Numerical Examples}
\label{sec:numex}

\subsection{Reduced Basis for a Fibrous Microstructure}
\label{sec:numex:fibre}
The applicability of the proposed RB method in combination with the sampling scheme is now numerically studied for a fibrous microstructure  roughly resembling polymers with woven reinforcements.
The goal is to prove the efficiency of the $\fF$-RB scheme in principle and under ``worst-case'' conditions, the latter meaning that the phase contrast is chosen to be rather large.
Yet, at this stage, it is explicitly not aspired to provide fully realistic examples. These would require investigations on the proper size of the microstructure and should employ dissipative material laws, both of which are not within the scope of this work.

A cubical microstructure with two fibrous inclusions is considered, see Figure~\ref{fig:numex:fibres:microstructure}a and cf.~\citep{kim2003} for a related example. The inclusions are periodic and occupy approximately 0.7\% of the volume. The mesh contains $35,516$ nodes in $25,633$ quadratic tetrahedron \color{black} elements (C3D10).

\def\microstructureheight{3.5cm}
\begin{figure}[H]
 \centering
 \captionbox*{(a)}[.25\textwidth][c]{\includegraphics[height=\microstructureheight]{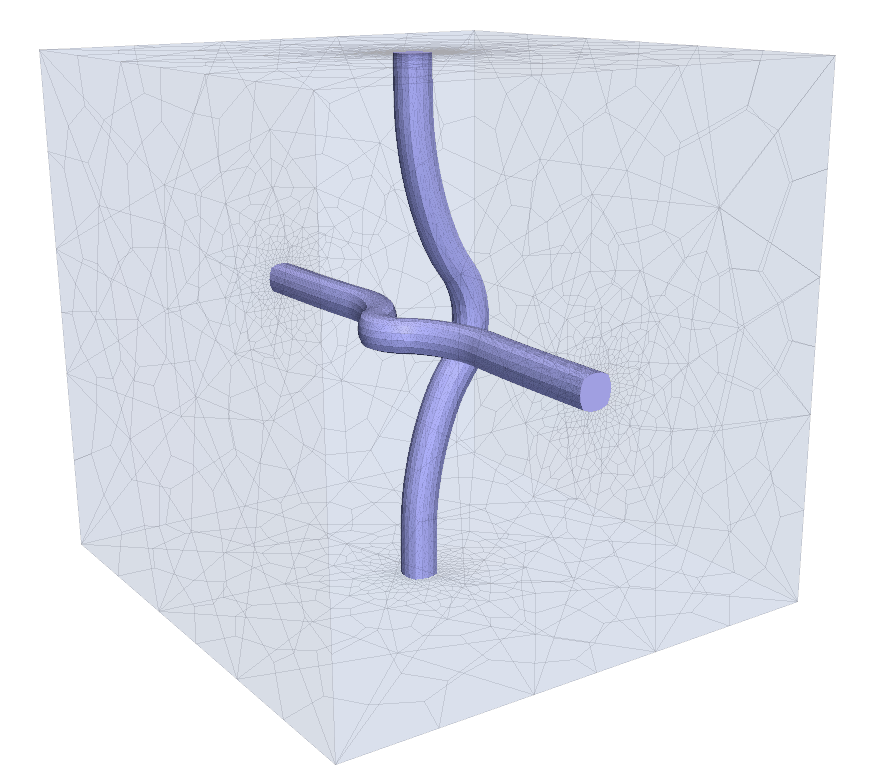}}%
 \captionbox*{(b)}[.25\textwidth][c]{\includegraphics[height=\microstructureheight]{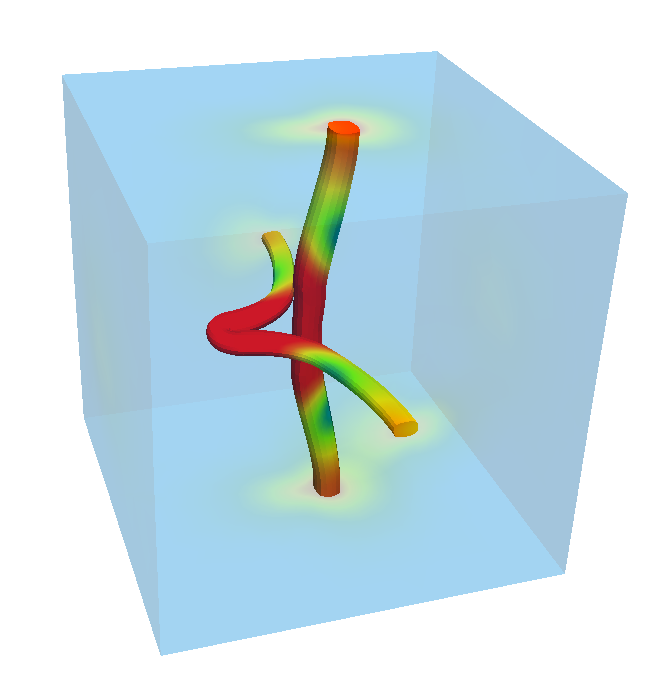} }%
 \captionbox*{(c)}[.25\textwidth][c]{\includegraphics[height=\microstructureheight]{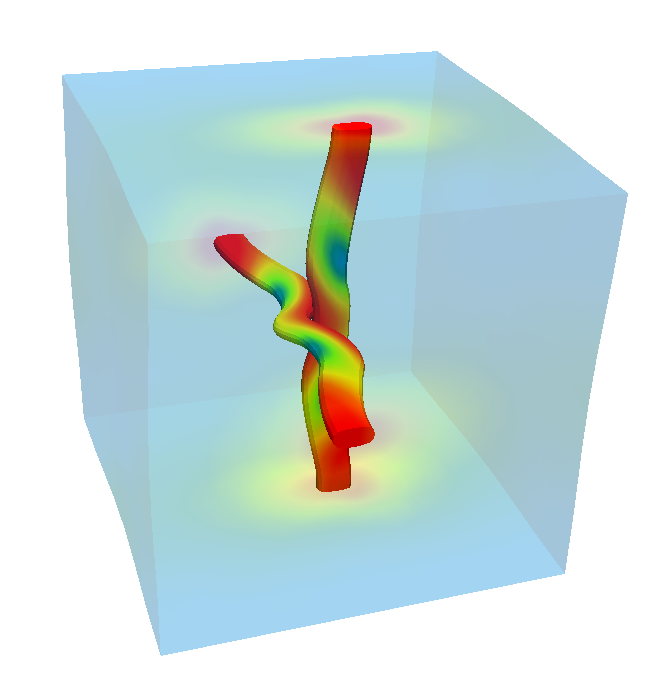}}%
 \captionbox*{(d)}[.25\textwidth][c]{\includegraphics[height=\microstructureheight]{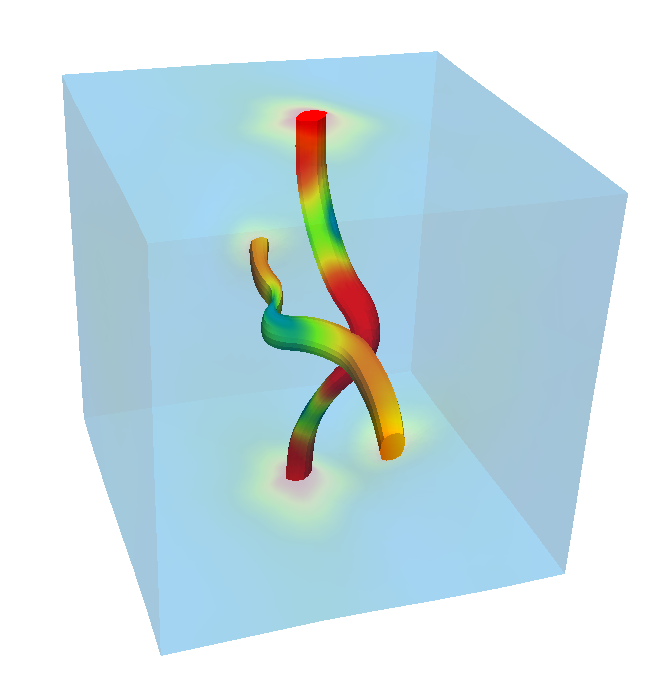}}%
 \caption{{(\textbf{a})} fibrous microstructure. {(\textbf{b})--(\textbf{d})} random $\fF$-reduced basis (RB) elements.}
 \label{fig:numex:fibres:microstructure}
\end{figure}

For the matrix, the material constants are chosen to be ${K_{\rm m}=400}$~MPa and ${G_{\rm m}=0.4}$~MPa, resembling rubber-like material properties. For the fibers, the values are ${K_{\rm f}=800}$~MPa and ${G_{\rm f}=240}$~MPa. The latter parameters approximate the behavior of stiffer polymers, such as polyethylene. The phase contrast is 600 with respect to the shear moduli, and the Poisson ratios are ${\nu_{\rm m} = 0.4995}$ and ${\nu_{\rm f}=0.3636}$. 

The training boundary conditions are defined with $N_{\rm dir}=128$ deviatoric directions, $\underline{N}\bn$, and $N_{\rm amp}=10$ deviatoric amplitudes, $t\bp$, which are regularly distributed in the interval $[0.05,0.5]$.
In order to also consider response to volumetric extension in the training data, an additional  set of $N_{\rm det}$=10 \color{black} boundary conditions of the form $\big(\ol{J}\bm\big)^{1/3}\fI$ is included in the training set, with the determinant $\ol{J}\bm$ being linearly increased from 1 to 1.02.

 The validation load cases are 640 mixed dilatational-deviatoric
 boundary conditions.
Along $N_{\rm dir}=64$ new deviatoric directions, both the deviatoric amplitudes ($t\bp=0.05,\ldots,0.5$) and the dilatational amplitudes ($\ol{J}\bm=0.9995,\ldots,0.995$) are applied in 10 equidistant increments.\color{black}

\begin{figure}[H]\centering
 \includegraphics[width=\textwidth]{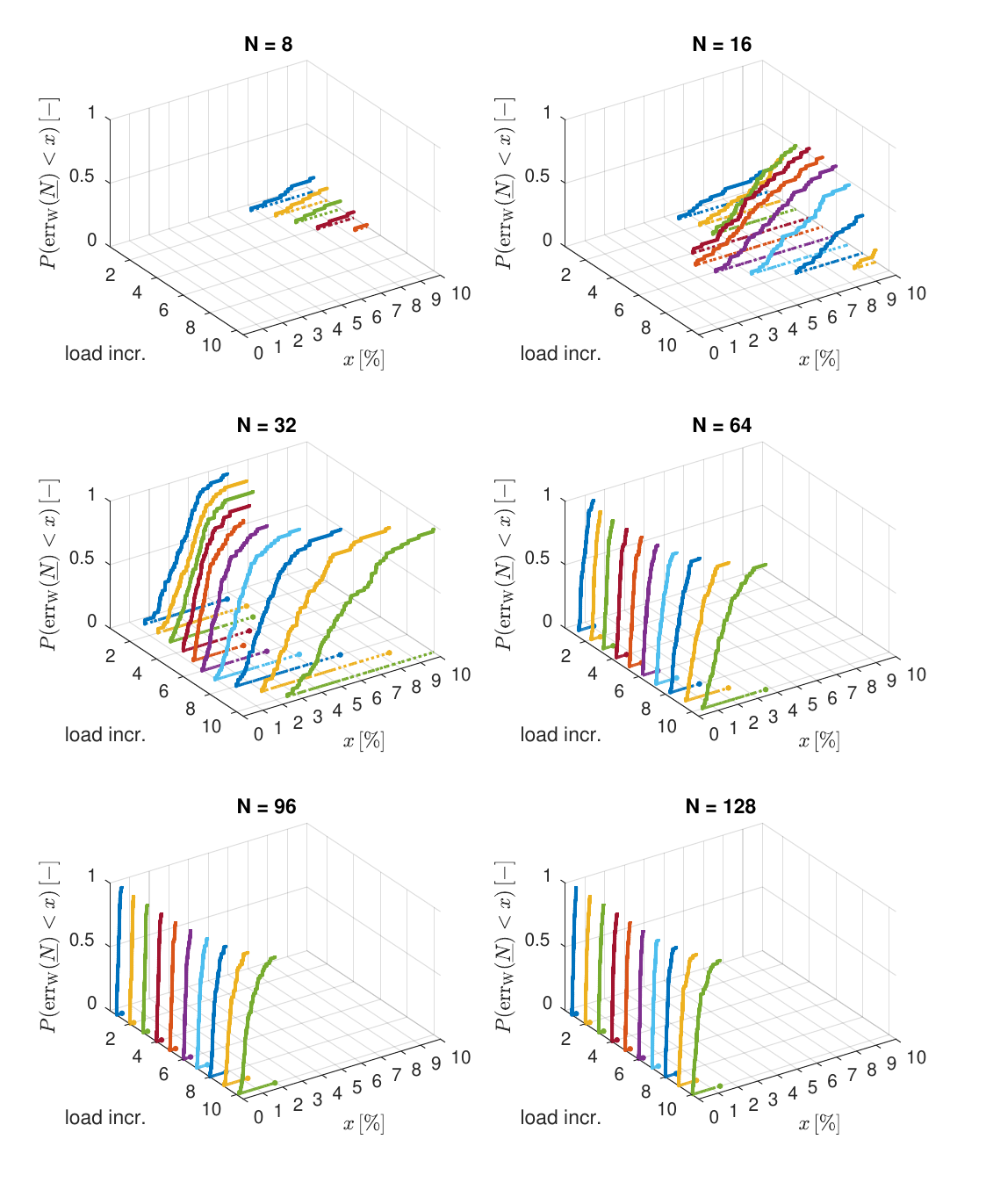}
 \caption{Cumulative energy error distribution per direction for the \emph{RB of the fibrous microstructure} under validation boundary conditions.}
 \label{fig:numex:fibres:W}
\end{figure}

\begin{figure}[H]\centering
 \includegraphics[width=\textwidth]{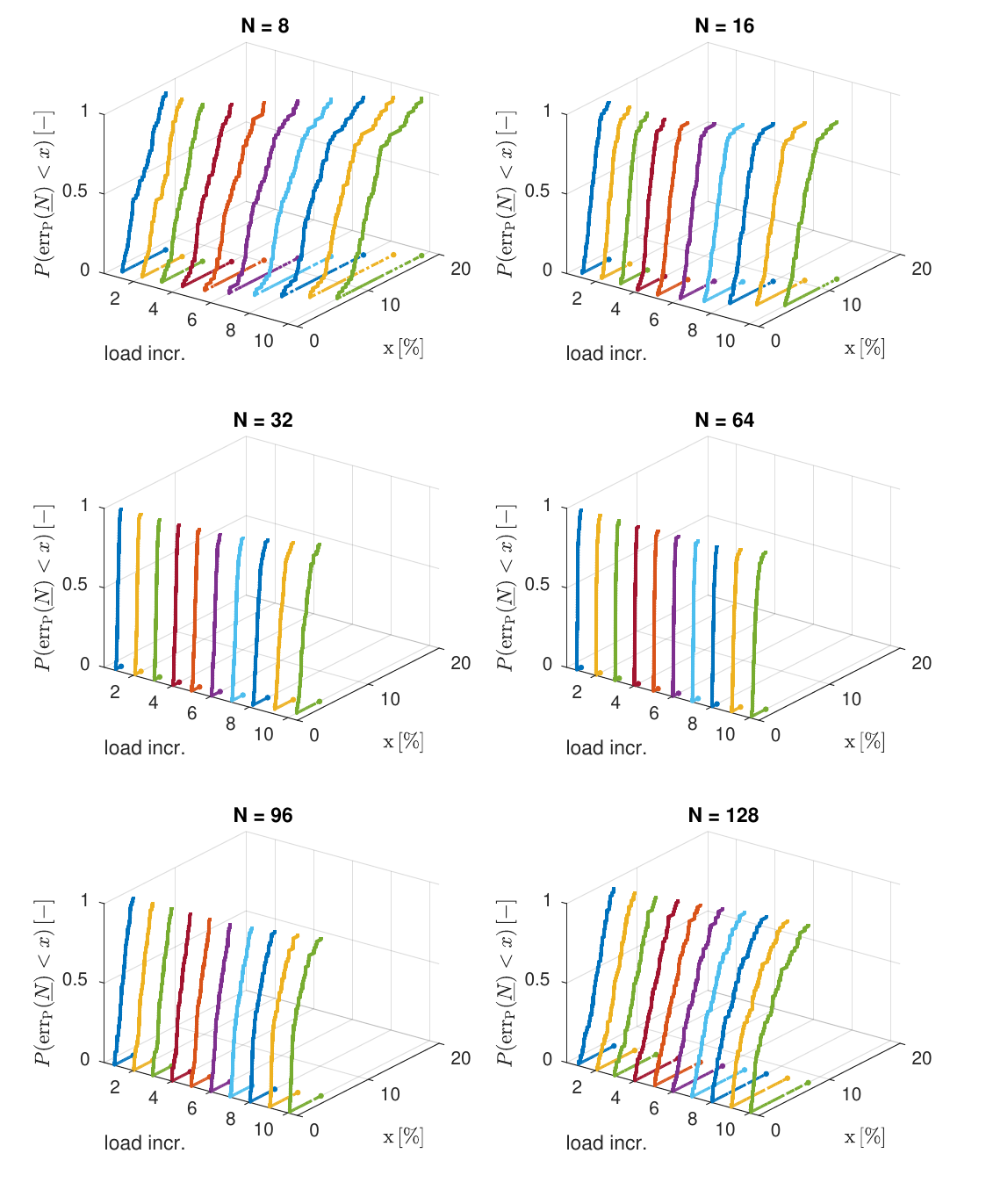}
 \caption{Cumulative stress error distribution per direction for the \emph{RB of the fibrous microstructure} under validation boundary conditions.}
 \label{fig:numex:fibres:F}
\end{figure}

The results for various values $N$ of the RB-size are compared with the results of FE simulations with the same boundary conditions. To this end, the error measures
\begin{align}\label{eq:numex:fibre:error}
 {\rm err}_{\rm W} &= \frac{\lVert \olW\RB - \olW^{\rm FEM}\rVert}{\lVert \olW^{\rm FEM}\rVert} \text{\rlap{$\quad\qquad$and}} & {\rm err}_{\rm P} &= \frac{\lVert \olfP\RB - \olfP^{\rm FEM}\rVert}{\lVert \olfP^{\rm FEM} \rVert}
\end{align}
are employed. Figures~\ref{fig:numex:fibres:W} and~\ref{fig:numex:fibres:F} visualize the results.

The distribution of the energy error, ${\rm err}_{\rm W}$, improves monotonically as the RB is enriched from $N=8$ to $N=128$ elements. This enrichment corresponds to the inclusion of additional subtrahends in the computation of $\olffC\RB$, improving the spectral over-estimation by the volume average of the stiffness, cf.~\eqref{eq:intro:micro:estimate_stiffness}. It is also noteworthy that the error tends to be higher for larger magnitudes of the applied kinematic boundary condition, although that is not always the case.

In contrast to this, the stress error ${\rm err}_{\rm P}$ distribution monotonically improves only up to a certain threshold value of the number of basis elements, which is $N=64$ in this example. For the greater bases with $N=96$ and $N=128$, the quality of the results deteriorates as far as the stress error is concerned. This is most likely due to an excessively oscillatory nature of the higher order modes---at some critical level $1\ll i=N_{\rm crit}<N$, the POD constructs eigenvectors $\underline{E}\bi$ with the $L^2(\varOmega_0)$-norm $\sqrt{\lambda_i}\ll1$. Therefore, the POD would construct basis vectors out of \emph{numerical fluctuations}, which would be unphysical. While the enrichment of the optimization space with unphysical information cannot increase the minimum energy error $\rm err_{\rm W}$, it might lead to fluctuations in the displacement field that have significant impact on the overall stress response. This is especially the case for numerical fluctuations within the stiff inclusion phase where low overall displacement errors still could lead to notable impact on the effective stress.

Nonetheless, all observed errors are less than 20\% and stay below 3\% for the optimal sampled size $N=64$. For half the basis size, $N=32$, the errors max at approximately 5\%, which is still acceptable considering the uncertainties involved in realistic two-scale simulations. Note that the maximum errors strongly depend on the maximum load amplitude, which is chosen to be very large in this example (50\% deviatoric strain and 0.5\% compression).

\begin{figure}[H]\centering
 \begin{minipage}{5cm}
  \includegraphics[scale=1]{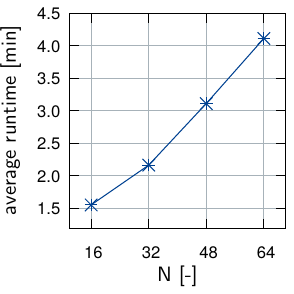}
 \end{minipage}%
  \caption{Laptop computer runtimes of the RB model for the fibrous microstructure for various sizes $N$ of the basis. Each data point represents the time needed for all 10 load increments. The spread of the individual times of the 64 validation cases around these average values is negligible.}
 \label{fig:numex:fibres:runtime}
\end{figure}

The runtimes of the RB model for different sizes $N$ are depicted in Figure~\ref{fig:numex:fibres:runtime}. A nearly linear growth of the runtime with respect to the basis size can be asserted. It is noteworthy that the online time of the RB method is strongly dominated by the assembly of the Jacobian $\ull{D}$. Therefore, a \emph{Quasi-Newton implementation} was chosen, resulting in only two assemblies per load increment.

Speed-ups become impressive when very large load increments are considered. In all examples observed thus far, \emph{the RB converges to the final load amplitude in a single increment, requiring 7--13 Quasi--Newton iterations, with only 2--4 assemblies of the Jacobi matrix $\ull{D}$ and a runtime of 10--50 seconds.}
This is in strong contrast to the FEM which is very sensitive to large load increments as they come along with a high probability of a violation of the condition ${\rm det}(\fF)>0$. By means of standard implementation, such occurrences are usually treated with cutbacks of the load increment, which is detrimental to the runtime of the FEM.

No rigorous speed-up analysis is intended at this point. Both the codes of the FEM and of the RB method are fairly efficient in-house C/C++ developments and perform exact line searches. While the FEM has not yet been equipped with a Quasi--Newton procedure, the linear solver makes use of parallelization. This is in contrary to the current implementation of the RB method. Depending on the microstructure (especially the geometry, material nonlinearities, and phase contrast), the loading conditions, and the size $N$ of the RB, \emph{speed-up factors are in the order of 5--100}.

\subsection{Reduced Basis for a Stiffening Microstructure}
\label{sec:numex:stiffening}
\begin{figure}[H]
 \centering
 \captionbox*{(a)}[.33\textwidth][c]{\includegraphics[height=\microstructureheight]{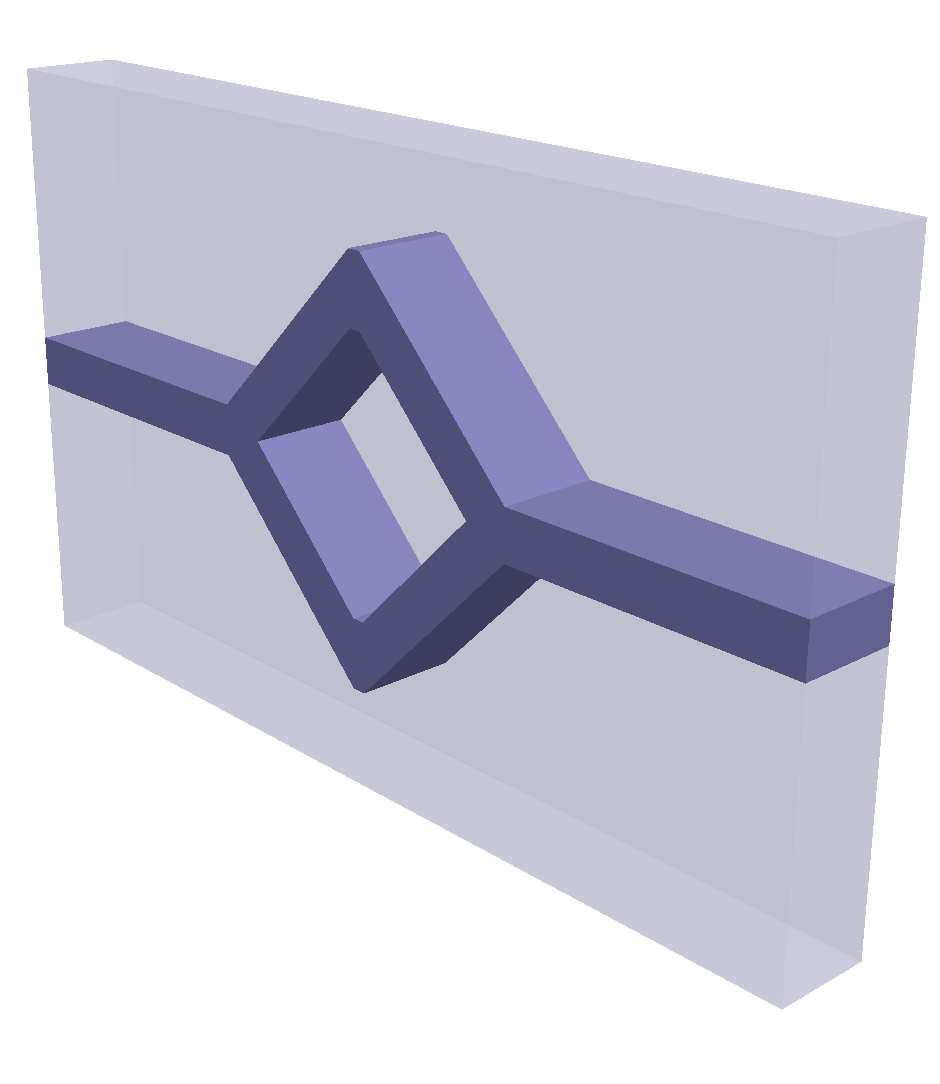}}%
 \captionbox*{(b)}[.33\textwidth][c]{\includegraphics[width=.33\textwidth]{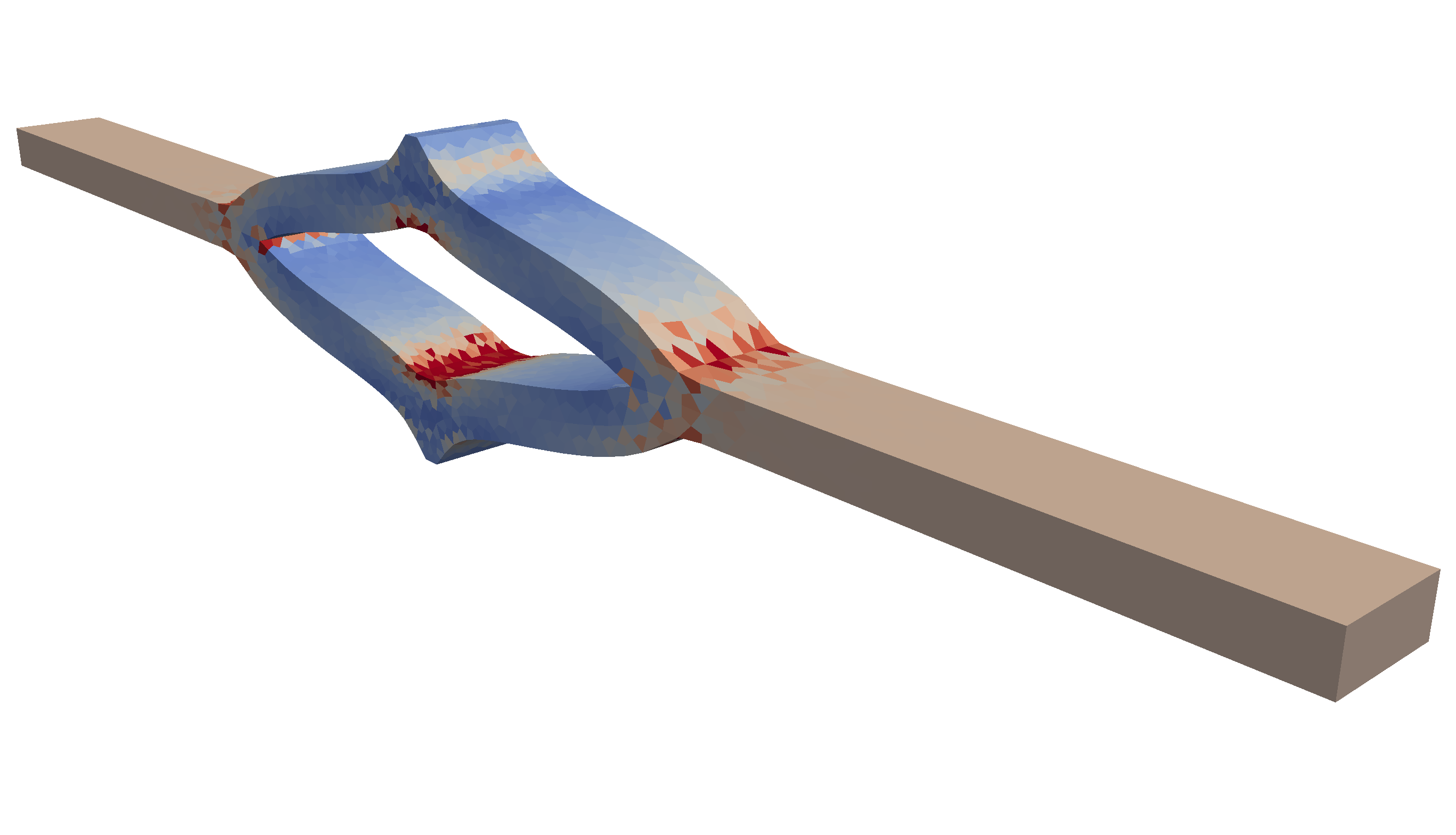} }%
 \captionbox*{(c)}[.33\textwidth][c]{\includegraphics[height=\microstructureheight]{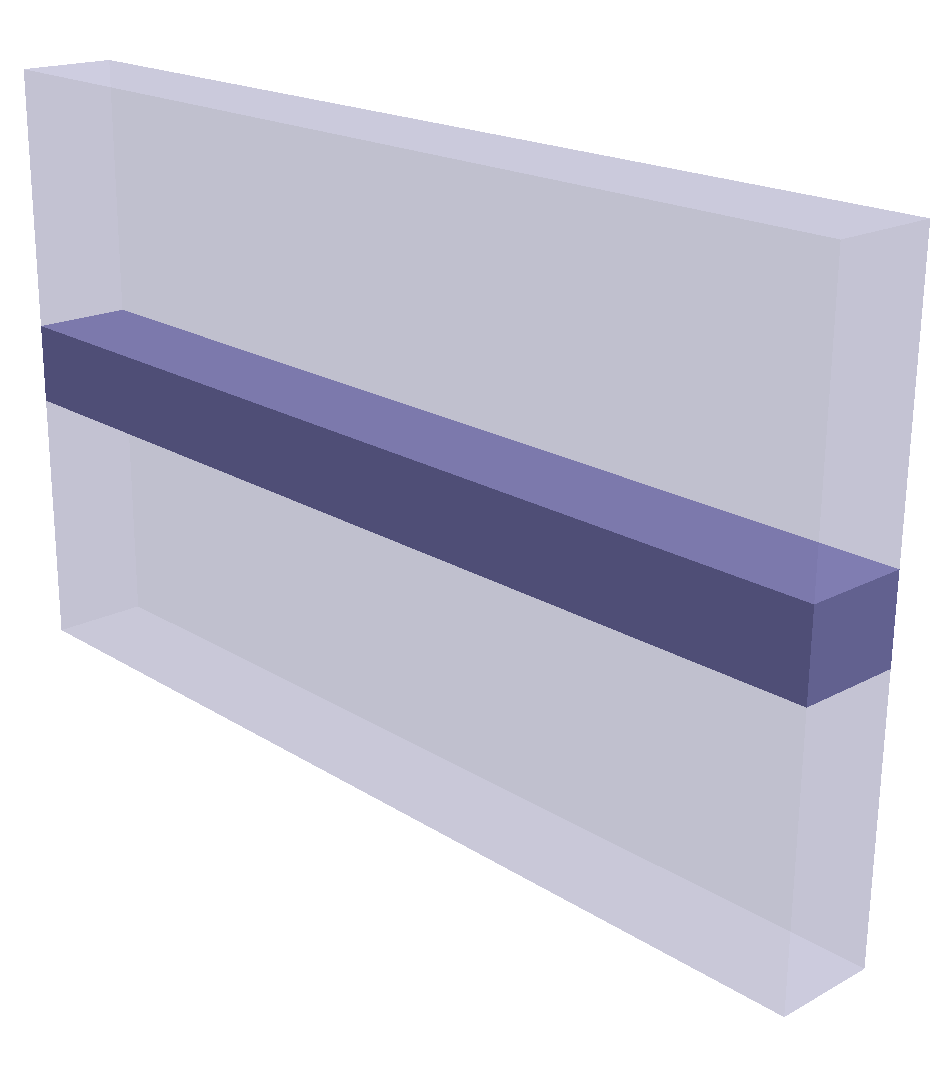}}%
 \caption{{(\textbf{a})} Cuboidal microstructure with hashlike inclusion phase. {(\textbf{b})} Deformed state under uniaxial tension loading. Only inclusion is shown, coloring indicates $\olfP_{xx}$. {(\textbf{c})} Straight inclusion substitute microstructure, leading to a comparable effective stress.}
 \label{fig:numex:stiffening:microstructure}
\end{figure}

The second example takes the ``worst-case'' approach further by considering a noncubical microstructure with even higher phase contrast and significant topological nonlinearity.
To this end, a cuboid microstructure occupying the domain $[-0.5,0.5]\times[-0.3,0.3]\times[-0.05,0.05]\subset\ffR^3$ and containing a hash-like inclusion is investigated, see Figure~\ref{fig:numex:stiffening:microstructure}a. The mesh is periodic and contains $33,923$ nodes in $21,726$ quadratic  tetrahedron \color{black} elements (C3D10). The reinforcement makes up approximately 13.3\% of the volume. Due to this large volume fraction, a pronounced geometry-induced nonlinearity of the effective response is expected under uniaxial loading conditions along the $x$-axis. As it is elongated, the hashlike part is straightened and thus increasingly aligned with the external loading, see Figure~\ref{fig:numex:stiffening:microstructure}b. Such effects might be desirable when designing microstructures for functional materials.

The material parameters are ${K_{\rm m}=19.867}$~MPa, ${G_{\rm m}=0.4}$~MPa, ${K_{\rm f}=19,867}$~MPa, and ${G_{\rm f}=400}$~MPa, implying a Poisson ratio of $0.49$ in both materials and a phase contrast of $1000$.

The training boundary conditions are the deviatoric ones of the set considered in Section~\ref{sec:numex:fibre}, i.e., $N_{\rm dir}=128$ deviatoric directions and $N_{\rm amp}=10$ regularly spaced deviatoric amplitudes from the interval $[0.05,0.5]$. No dilatational training cases are considered, i.e., only points from a five-dimensional submanifold of the space $\ol\cU$ are sampled.

Uniaxial tension boundary conditions are applied for the validation. More precisely, the stretch component $\olfU_{\rm xx}$ is increased from $1.0$ to $1.5$ in $10$ increments of equal size. The other components are chosen such that all but the $\rm xx$-component of the effective stress $\olfP$ vanish.

\begin{figure}[H]\centering
 \begin{minipage}{.42\textwidth}
  \includegraphics[scale=1]{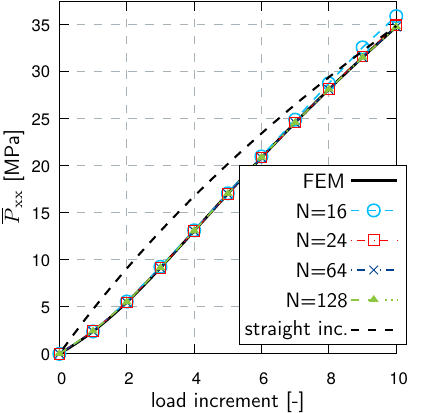}
 \end{minipage}%
  \caption{Stress curves for a microstructure with geometric stiffening (cf. Figure~\ref{fig:numex:stiffening:microstructure}a), comparing the FEM result to the RB for various number of basis elements $N$. A similar microstructure without geometric stiffening but with the same final stress value (cf. Figure~\ref{fig:numex:stiffening:microstructure}c) leads to the black dashed curve.}
  \label{fig:numex:stiffening:res}
\end{figure}

Figure~\ref{fig:numex:stiffening:res} depicts the results for different sizes $N$ of the RB. The influence of the stiffening effect on the stress curve is emphasized by the black dashed line corresponding to a similar microstructure with a straight, cuboid inclusion that leads to the same final stress value under these boundary conditions, see Figure~\ref{fig:numex:stiffening:microstructure}c.

In this example, the geometric stiffening effect is captured by the RB with high accuracy, with as few as $N=24$ basis elements. For moderate stretches, even an RB size of $N=16$ suffices. These results are somewhat more impressive when noticing that the applied boundary condition contains more than 1.2\% volumetric compression, i.e. the validation loading actually lies outside the submanifold covered during training.

\begin{figure}[H]\centering
 \includegraphics[scale=0.9]{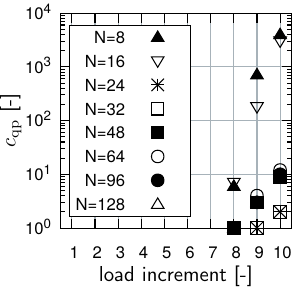}$\quad$%
 \includegraphics[scale=0.9]{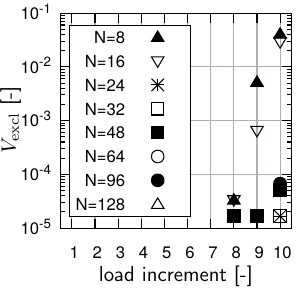}
 \caption{{(Left)} Number of quadrature points at which the cutoff function~$\phi$ attains a value less than one. {(Right)} Relative excluded volume.}
 \label{fig:numex:stiffening:phi}
\end{figure}

In order to examine the action of the cutoff function $\phi$, the following two indicators are introduced:
\begin{align}
 \label{eq:numex:stiffening:cqp}c_{\rm qp} &= \text{\# quadrature points with } \lb\phi(J)<1\rb,\\
 \label{eq:numex:stiffening:Vexcl}V_{\rm excl} &= \lb\lvert\varOmega_0\rvert-\sum_{p=1}^{N_{\rm qp}}\phi(J_p)w_p\rb/\lvert\varOmega_0\rvert.
\end{align}
The first quantity, $c_{\rm qp}$, counts the number of quadrature points at which the cutoff function has an influence. The second one, $V_{\rm excl}$, measures the relative excluded volume, interpreting the value of~$\phi$ as a scaling of the corresponding quadrature weight. The values of these indicators are depicted in Figure~\ref{fig:numex:stiffening:phi}.

Most notably, the cutoff function does not have any effect before the eighth load increment in this example. Only for large load amplitudes does this numerical stability tweak become necessary. Even then, the number of points at which it has an influence is small, considering the total number of quadrature points, $N_{\rm qp}=86,904$. This example is representative for all conducted numerical experiments.


\section{Discussion}
\label{sec:discussion}

\subsection{Discussion of the Reduced Basis Method}
\label{sec:discussion:RB}

\subsubsection{Relation of the RB Homogenization to Analytical Estimates}
Zero coefficients, $\underline{\xi}=\underline{0}$, correspond to the Taylor homogenization, i.e., to the nonlinear counterpart of the Voigt estimate \citep{voigt10}, which provides an upper bound for the material response, cf.~\eqref{eq:intro:micro:estimate_stiffness}. Starting with the initial guess $\underline{\xi}^{(0)} = \underline{0}$, the evolution of the coefficients corresponds to a  (possibly not monotonous) \color{black} relaxation of this overly stiff response into a more natural state. In view of improved computational efficiency, a  nonzero \color{black} initial guess $\underline{\xi}^{(0)}$ combined with an exact line search has proven reasonable and easy to realize. For instance, such a guess might stem from previous load steps or an interpolation/extrapolation of available coefficient data.

\subsubsection{Reconstruction of Displacement Fields}
Given an RB approximation of the deformation gradient, $\fF\RB$, one can reconstruct the corresponding displacement field uniquely up to rigid body motion. This is possible due to the linear dependence of the deformation gradient fluctuations on the displacement fluctuations. Recall the definition of the RB in~\eqref{eq:RB:identification:definiton},
\begin{align}\notag
 \fB\bi &= \sum_{j=1}^{N_{\rm s}} \frac{1}{\sqrt{\lambda_i}}\lb\underline{E}\bi\rb_j \widetilde{\fF}\bj \quad (i=1,\ldots,N).
\intertext{The corresponding displacement fluctuations are}
\label{eq:resume:reconstruction:uB}
 \widetilde{\fu}_{\rm B}\bi &= \sum_{j=1}^{N_s} \frac{1}{\sqrt{\lambda_i}}\lb\underline{E}\bi\rb_j \widetilde{\fu}\bj \quad (i=1,\ldots,N).
\intertext{The displacement fluctuation fields $\widetilde{\fu}\bj$ are defined by $\widetilde{\fu}\bj(\fX)=\fu\bj(\fX)-{\ol\fH}\bj\fX$, where the displacement fields $\fu\bj$ are the solutions computed during training, and ${\ol\fH}\bj={\ol\fU}\bj-\fI$. Thus, a displacement field compatible to the RB result $\fF\RB(\fX;\olfF)$ is given by} 
\label{eq:resume:reconstruction:u}
 \fu\RB(\fX;\olfF) &= \ol\fH\fX+\sum_{i=1}^N\xi^*_i(\olfF)\widetilde{\fu}_{\rm B}\bi(\fX).
\end{align}
The missing term $\ol\fu(\ol\fX)$, cf.~\eqref{eq:intro:micro:displacement}, cannot be reconstructed.

\subsubsection{Relation to Classical Displacement-Based POD Methods}
In a certain sense, the entries of the correlation matrix used in the offline phase, cf. Section~\ref{sec:RB:identification}, are \emph{``weighted'' scalar products} of the displacement fluctuation fields $\widetilde{\fw}\bi$ within the Sobolev space~$H^1_0(\varOmega_0)$. ``Weighted'' is to be understood in that the zeroth order derivative is multiplied by zero. Classical displacement-based POD methods compute correlations of the fluctuations $\widetilde{\fw}\bi$ within the Lebesgue space~$L^2(\varOmega_0)$. The change to $H^1_0(\varOmega_0)$-like scalar products is physically motivated by the fact that the local energy $W=W(\fF)$ explicitly depends on the gradient of the displacement, $\fF=\fu\otimes\nabla_{\rm X}+\fI$, but \emph{not} on the displacement, $\fu$.

\subsubsection{Advantages Compared to General Displacement-Based Schemes}
The proposed method is advantageous compared to both displacement-based POD methods and the classical FEM for the following reasons:
\begin{itemize}
 \item \emph{No gradients need to be computed from displacement fields}, which displacement-based schemes always require prior to the evaluation of the material law.
 \item The residual $\underline{r}$ and the Jacobian $\ull{D}$ are \emph{algorithmically sleek and trivial to implement}.
 \item  The \emph{absence of element formulations} in the assembly of the reduced residual~$\underline{r}$ and of the Jacobi matrix~$\ull{D}$ contributes to both the simplicity and the efficiency of the method---no incidence matrices occur, allowing for \emph{linear memory access}. Moreover, the algebraic operations associated with reference element formulations are bypassed. This is also in favor of \emph{parallel computations}. Such an implementation is still outstanding for the problem at hand, but has been conducted for related problems in the small strain setting in~\citep{fritzen2016b}.\color{black}
\end{itemize}

Although the storage of the basis $\{\fB\bi\}_{i=1}^N$ requires $9N_{\rm qp}N$ double precision values, the basis is compact enough to be completely loaded into random access memory of standard computers\color{black}. For instance, the bases considered in Section~\ref{sec:numex} occupy only \textasciitilde200~Mb of memory for $N=32$.

We now briefly address the \emph{algorithmic complexity} associated with the proposed $\fF$-RB method and with the $\fu$-RB method that was employed in previous works, such as \citep{radermacher2016} and~\citep{yvonnet2007}. To this end, the fully discretized versions of the residual $\underline{r}$ and of the Jacobian $\ull{D}$ as well as discrete quantities associated with the $\fu$-RB method are introduced in the following listing.

\begin{itemize}
 \item $\underline{P}(\fX_p)\in\ffR^{9}$: Nine values of the stress $\fP(\fF_\xi(\fX_p))$ at the quadrature point $\fX_p\in\varOmega_0$
 \item $\ull{C}(\fX_p)\in\ffR^{9\times9}$: Symmetric stiffness tensor 
 \item $\ull{B}(\fX_p)\in\ffR^{9\times N}$: $\fF$-RB matrix containing the nine values of each basis elements $\fB\bi$ as columns
 \item $w_p$: The quadrature weight at $\fX_p$
 \item $N_{\rm DOF}$: Three times the number of nodes, $\sim N_{\rm qp}$ 
 \item $\underline{r}^{\,\rm FE}\in\ffR^{N_{\rm DOF}}$: Global FE residual vector
 \item $\ull{B}^{\rm u}\in\ffR^{N_{\rm DOF}\times N}$: $\fu$-RB matrix of which the columns contain the nodal displacement values\color{black}
 \item $\ull{K}^{\,\rm FE}\in\ffR^{N_{\rm DOF}\times N_{\rm DOF}}$: Global FE stiffness matrix
\end{itemize}

Table~\ref{tab:complexity} compares the \emph{algorithmic complexity} of the presented $\fF$-RB method with that of standard $\fu$-RB schemes. First of all, both methods share a \emph{quadratic} dependence of their Jacobi matrices on the number of modes, $N$. Therefore, the assembly of the Jacobian is usually the most costly operation. Secondly, both approaches' complexities suffer a linear dependency on the number of quadrature points, $N_{\rm qp}$. In the displacement-based approach, this is included in the assembly of the residual and of the stiffness, which relate to the factor 9 and 9\textsuperscript{2}, respectively. Thirdly, the novel $\fF$-RB scheme spares the computational overhead associated with FE formulations $\underline{r}^{\rm FE}$ and $\ull{K}^{\rm FE}$.  More details on this matter are currently being investigated.\color{black}

\begin{table}[H]
\centering
 \caption{Algorithmic complexities of the well-established $\fu$-based RB method and the novel $\fF$-based RB method. The assembly of the FE residual and of the FE stiffness depend on $N_{\rm qp}$.}
 \label{tab:complexity}
\begin{tabular}{ccc}
\toprule
 \textbf{RB Method} & \textbf{Quantity} & \textbf{Complexity}\\
 \midrule
 \multirow{2}{*}[-.35cm]{$\fF$-based} &\rule[-.3cm]{0cm}{.5cm} $\displaystyle \underline{r} = \sum_{p=1}^{N_{\rm qp}}\ull{B}(\fX_p)\TT\underline{P}(\fX_p)w_p$ & $\cO(9NN_{\rm qp})$\\
 \cline{2-3} &\rule{0pt}{.7cm} $\displaystyle \ull{D} = \sum_{p=1}^{N_{\rm qp}}\ull{B}(\fX_p)\TT\ull{C}(\fX_p)\ull{B}(\fX_p)w_p$ & $\cO([9^2N+9N^2]N_{\rm qp})$\\
 \midrule
 \multirow{2}{*}[-.07cm]{$\fu$-based} &\rule[-.2cm]{0pt}{0pt} $\displaystyle\underline{r}=\lb\ull{B}^{\rm u}\rb\TT\underline{r}^{\rm FE}$ & $\cO(NN_{\rm DOF})$ + assembly of $\underline{r}^{\rm FE}$\\
 \cline{2-3}&\rule{0pt}{.47cm} $\displaystyle\ull{D}=\lb\ull{B}^{\rm u}\rb\TT\ull{K}^{\rm FE}\ull{B}^{\rm u}$ & $\cO( NN_{\rm DOF}^2+\color{black}N^2N_{\rm DOF})$ + assembly of $\ull{K}^{\rm FE}$\\
\bottomrule
\end{tabular}
\end{table}

\subsubsection{Outlook}
Future research should aim at an application of the introduced Reduced Basis method within realistic two-scale simulations, in analog to \citep{Fritzen2018,fritzen2016b,rambausek2019,kochmann2017}. Hyperreduction methods, cf. \cite{Ryckelynck2005}, might give rise to additional speed-ups in the online phase. Further, modifications of the cutoff function, $\phi$, should be investigated---a function with compact support might be more appropriate.
The construction of the RB from large sets of snapshots is computationally intense, as much data needs to be processed. In the above examples, the POD consumed multiple hours of time. Hierarchical approximations, such as~\cite{himpe2018}, might mitigate the effects by enabling parallel computations. Overall, the long-term perspective is to extend this RB framework efficiently to the context of dissipative materials. \color{black}

\subsection{Discussion of the Sampling Strategy}
\label{sec:discussion:sampling}
The proposed sampling strategy is meant to serve as a template. As exemplified in Section~\ref{sec:numex:fibre}, the samplings can be modified and still lead to a coverage of the set of macroscopic boundary conditions that is sufficient for the problem at hand. The example of Section~\ref{sec:numex:stiffening} took this idea further and showed that it might not even be necessary to sample the macroscopic determinant. Hence, the sampling can sometimes be reduced to the five-dimensional subspace of isochoric macroscopic stretch tensors.

In any case, the exact choice of both the inputs to Algorithm~\ref{algo:sampling} and the distributions of the deviatoric amplitudes and the macroscopic determinants remains to be based on knowledge and sophisticated guesses, at least at the current state of the art. Further research on this matter might lead to a refined alternative to Algorithm~\ref{algo:sampling}, possibly involving the evaluation of error estimators.

\vspace{6pt} 



\authorcontributions{
    Conceptualization, Oliver Kunc and Felix Fritzen; Funding acquisition, Felix Fritzen; Methodology, Oliver Kunc and Felix Fritzen; Project administration, Felix Fritzen; Software, Oliver Kunc; Supervision, Felix Fritzen; Writing---original draft, Oliver Kunc and Felix Fritzen.
} 

\funding{This research was funded by Deutsche Forschungsgemeinschaft (DFG) within the Emmy--Noether programm under grant DFG-FR2702/6.}


\conflictsofinterest{The authors declare no conflict of interest.} 

\abbreviations{The following abbreviations are used in this manuscript:\\

\noindent 
\begin{tabular}{@{}ll}
RB & Reduced Basis\\
FE(M) & Finite Element (Method)\\
POD & Proper Orthogonal Decomposition\\
DOF & degree(s) of freedom\\
FOM & full-order model\\
s.p.d. & symmetric positive definite \\
DDMS & Dilatational-Deviatoric Multiplicative Split
\end{tabular}}

\appendixtitles{no} 
\appendix
\section{Material Objectivity}
\label{app:objectivity}
The components of the stiffness tensor $\ffC$ show the following transformation behavior. 
 \begin{align}\label{eq:app:objectivity:ffC}
  C_{ijkl}(\fF)=\pd{P_{ij}(\fF)}{F_{kl}} &\overset{\eqref{eq:sampling:objectivity:P}}{=} \sum_{m=1}^3\pd{R_{im}P_{mj}(\fU)}{F_{kl}} = \sum_{m=1}^3 \left(\pd{R_{im}}{F_{kl}}P_{mj}(\fU) + R_{im}\pd{P_{mj}(\fU)}{F_{kl}}\right)\\
  &\overset{\phantom{\eqref{eq:sampling:objectivity:P}}}{=} \sum_{m,n,o=1}^3 \left(\delta_{ik}U^{-1}_{lm}P_{mj}(\fU) + R_{im}\pd{P_{mj}(\fU)}{U_{no}}\pd{U_{no}}{F_{kl}}\right)\\
  &\overset{\phantom{\eqref{eq:sampling:objectivity:P}}}{=} \sum_{m,n,o=1}^3 \left(\delta_{ik}U^{-1}_{lm}P_{mj}(\fU) + R_{im}C_{mjno}(\fU)R_{kn}\delta_{ol}\right)\\
  &\overset{\phantom{\eqref{eq:sampling:objectivity:P}}}{=} \sum_{m,n=1}^3 \left(\delta_{ik}U^{-1}_{lm}P_{mj}(\fU) + R_{im}C_{mjnl}(\fU)R_{kn}\right)
 \end{align}\color{black}
Here, $\delta_{ij}$ is the Kronecker symbol, and $i,j,k,l=1,2,3$.
\section{Effective Material Responses of the RB}
\label{app:effective}
Let $\ffI$ denote the fourth order identity tensor and let the arguments of the $\fF$-RB approximation \eqref{eq:RB:ansatz} be omitted, i.e. here $\fF\RB=\fF\RB(\fX;\olfF)$. Its derivative with respect to the boundary condition $\olfF$ is
\begin{align}\label{eq:app:effective:aux}
 \pd{\fF\RB}{\olfF} &= \ffI + \sum_{i=1}^N\fB\bi\otimes\pd{\xi_i^*}{\olfF}(\olfF).
\end{align}

\subsection{Effective Stress}
\label{app:effective:stress} 
\begin{align}\label{eq:app:effective:stress}
 \begin{split}
    \olfP\RB  &\overset{\phantom{\eqref{eq:app:effective:aux}}}{=} \pd{\olW\RB}{\olfF}
                = \pd{}{\olfF}\left<W\RB\right>
                = \left< \pd{W\RB}{\fF}\cdot\pd{\fF\RB}{\olfF}\right> \\
                &\overset{\eqref{eq:app:effective:aux}}{=} \left< \fP\RB \right> + \sum_{i=1}^N\left< \fP\RB \cdot\lb\fB\bi\otimes\pd{\xi_i^*}{\olfF}(\olfF)\rb \right>\\
                &\overset{\phantom{\eqref{eq:app:effective:aux}}}{=}\left< \fP\RB \right> + \sum_{i=1}^N\left< \fP\RB \cdot\fB\bi\right>\otimes\pd{\xi_i^*}{\olfF}(\olfF) \qquad\quad\overset{\eqref{eq:RB:residual}}{=} \left< \fP\RB \right>
 \end{split}
\end{align}\color{black}

\subsection{Effective Stiffness}
\label{app:effective:stiffness}
\begin{align}\label{eq:app:effective:stiffness_1}
 \begin{split}
 \olffC\RB    &\overset{\phantom{\eqref{eq:app:effective:aux}}}{=} \pd{\olfP\RB}{\olfF}
                \overset{\eqref{eq:app:effective:stress}}{=} \pd{}{\olfF}\left<\fP\RB\right>
                = \left< \pd{{}^2W\RB}{\olfF\partial\fF} \right>
                = \left< \pdII{W\RB}{\fF}\cdot\pd{\fF\RB}{\olfF} \right> \\
                &\overset{\eqref{eq:app:effective:aux}}{=} \left< \ffC\RB \right> + \sum_{i=1}^N \left< \ffC\RB\cdot\lb\fB\bi\otimes\pd{\xi_i^*}{\olfF}(\olfF)\rb \right> \\
                &\overset{\phantom{\eqref{eq:app:effective:aux}}}{=} \left< \ffC\RB \right> + \sum_{i=1}^N \left< \ffC\RB\cdot\fB\bi\right>\otimes\pd{\xi_i^*}{\olfF}(\olfF)
 \end{split}
\end{align}
For $\pd{\xi_i^*}{\olfF}(\olfF)$, we demand that the residual $r_i(\olfF,\underline{\xi})$ from \eqref{eq:RB:residual} is stable with respect to the boundary condition $\olfF$ when converged to $r_i(\olfF,\underline{\xi}^*(\olfF))=0$,
\begin{align}\label{app:effective:residual}
 \begin{split}
  \pd{r_i}{\olfF}(\olfF,\underline{\xi}^*(\olfF)) &= \left< \fB\bi \cdot \pd{\fP\RB}{\olfF} \right> \\
  &=\left< \fB\bi\cdot\lb\pd{\fP\RB}{\fF}\pd{\fF\RB}{\olfF}\rb \right>\\
  &=\left<\fB\bi\cdot\ffC\RB \right>
    + \sum_{j=1}^N\left<\fB\bi\cdot\ffC\RB \cdot \pd{\fF\RB}{\xi_j^*}\pd{\xi_j^*}{\olfF}\right>\\
  &=\left<\fB\bi\cdot\ffC\RB \right>
    + \sum_{j=1}^N\underbrace{\left<\fB\bi\cdot\ffC\RB \cdot \fB\bj\right>}_{D_{ij}} \pd{\xi_j^*}{\olfF} \quad= \fzero.
 \end{split}
\end{align}
It follows that
\begin{align}\label{app:effective:residual:aux}
 \begin{split}
  \pd{\xi_j^*}{\olfF}(\olfF) &= - \sum_{i=1}^N \lb \ull{D}^{-1}\rb_{ij}\left<\fB\bi\cdot\ffC\RB\right>.
 \end{split}
\end{align}

\section{Basis for Symmetric Traceless Second Order Tensors}
\label{app:exp:basis}
\begin{align}\begin{split}\label{eq:sampling:basis}
 \ull{Y}^{(1)} &=\sqrt{\frac{1}{6}}\sV 2 &\pmo &\pmo \\ 0 &-1 &\pmo \\0&\pmo &-1\eV \qquad
 \ull{Y}^{(2)} =\sqrt{\frac{1}{2}}\sV 0 &\pmo &\pmo \\ 0 &\phantom{-}1 &\pmo \\0&\pmo &-1\eV \\
 \ull{Y}^{(3)} &=\sqrt{\frac{1}{2}}\sV 0 &\phantom{-}1 &\pmo \\ 1 &\pmo &\pmo \\0&\pmo &\pmo \eV \qquad
 \ull{Y}^{(4)} =\sqrt{\frac{1}{2}}\sV 0 &\pmo &\phantom{-}1  \\ 0 &\pmo &\pmo \\1&\pmo &\pmo \eV \qquad
 \ull{Y}^{(5)} =\sqrt{\frac{1}{2}}\sV 0 &\pmo &\pmo  \\ 0 &\pmo &\phantom{-}1 \\0&\phantom{-}1 &\pmo \eV 
\end{split}\end{align}


\reftitle{References}





\end{document}